\documentclass[BTech]{iitmdiss}

\usepackage{titlesec}
\usepackage{times}
\usepackage{t1enc}
\usepackage{float}
\usepackage{amsmath, amssymb}
\usepackage{mathrsfs}
\usepackage[autostyle]{csquotes}
\usepackage{longtable}
\usepackage{lipsum}
\usepackage{amsfonts}
\usepackage{amstext}
\usepackage{threeparttable}
\usepackage{textcomp}
\usepackage{rotating}
\usepackage{lscape}
\usepackage{gensymb}
\usepackage{placeins}
\usepackage{graphicx}
\usepackage{epstopdf}
\usepackage{wrapfig}
\usepackage{tabularx}
\usepackage{subfig}
\usepackage{multirow}
\usepackage{amsmath}
\usepackage{makecell}
\usepackage{hhline}
\usepackage{cellspace}
\usepackage{enumitem}
\usepackage{mwe}
\usepackage{hyperref}
\usepackage{setspace}
\usepackage{adjustbox}
\usepackage{ltablex} 
\usepackage{inputenc}
\usepackage{color}
\usepackage{framed}
\usepackage{afterpage}
\usepackage{array}
\usepackage{multirow}
\usepackage{geometry}
\usepackage{capt-of}
\usepackage{soul}
\usepackage{booktabs}
\usepackage{afterpage}
\usepackage{bm}
\usepackage[usenames,dvipsnames,svgnames,table]{xcolor} 
\usepackage{tikz}
\usetikzlibrary{shapes.geometric, arrows}
\usetikzlibrary{positioning}
\usepackage{verbatim}

\begin{document}

%%%%%%%%%%%%%%%%%%%%%%%%%%%%%%%%%%%%%%%%%%%%%%%%%%%%%%%%%%%%%%%%%%%%%%
% Title page

\title{Models for Characterizing Darrieus Turbines and Synthesizing Wind Speed Scenarios}

\author{ASHWATH MUKESH BHAT\\ME17B002}

\date{JUNE 2021}
\department{MECHANICAL ENGINEERING}

%\nocite{*}
\maketitle

%%%%%%%%%%%%%%%%%%%%%%%%%%%%%%%%%%%%%%%%%%%%%%%%%%%%%%%%%%%%%%%%%%%%%%
% Certificate
\certificate

\vspace*{0.5in}

\noindent This is to certify that the project titled {\bf Models for Characterizing Darrieus Turbines and Synthesizing Wind Speed Scenarios}, submitted by {\bf Ashwath Mukesh Bhat (ME17B002)}, to the Indian Institute of Technology, Madras, for the award of the degree of {\bf Bachelor of Technology}, is a bona fide record of the project work done by him under our supervision. The contents of this report, in full or in parts, have not been submitted to any other Institute or University for the award of any degree or diploma.

\vspace*{1.75in}

\begin{singlespacing}
\hspace*{-0.25in} 
\parbox{3.0in}{
\noindent {\bf Prof. Dhiman Chatterjee} \\
\noindent Professor \\
\noindent Dept. of Mechanical Engineering\\
\noindent Indian Institute of Technology Madras
} 
\hspace*{0.5in} 
\parbox{3.0in}{
\noindent {\bf Prof. Satyanarayanan Seshadri} \\
\noindent Associate Professor \\
\noindent Dept.  of  Applied Mechanics\\
\noindent Indian Institute of Technology Madras
}  
\end{singlespacing}

\vspace*{0.5in}
\noindent Place: Chennai - 600 036\\
Date: 18\textsuperscript{th} June 2021

%%%%%%%%%%%%%%%%%%%%%%%%%%%%%%%%%%%%%%%%%%%%%%%%%%%%%%%%%%%%%%%%%%%%%%
% Acknowledgements
\acknowledgements

I would like to express my sincere gratitude to Prof. Satya and Prof. Dhiman for their guidance through the project, and the support and advice received outside of this project. I am also thankful to the many Professors of the Applied Mechanics and Mechanical Engineering departments who have taught me over the past four years, and cemented my interest in engineering. Finally, I would like to express my thanks to all my friends from insti, who've made the past 4 years memorable, and my family for their unconditional love and support. 

%%%%%%%%%%%%%%%%%%%%%%%%%%%%%%%%%%%%%%%%%%%%%%%%%%%%%%%%%%%%%%%%%%%%%%
% Abstract

\abstract

\noindent KEYWORDS: \hspace*{0.5em} \parbox[t]{4.4in}{Darrieus Turbine; Double-Multiple Streamtube Method; OpenFOAM; 2-D CFD Simulation; Autoregressive Moving Average Model; Synthetic Wind Speed.}

\vspace*{24pt}

\noindent With the evolving global energy dynamics, new and innovative ways to integrate renewable energy technologies into residential and commercial complexes are being adopted. For this purpose, techno-economic models such as NREL-SAM and HOMER are handy for evaluating such integration of renewables. However, the unavailability of libraries for vertical axis wind turbines (VAWT) and wind speed data limitations have been identified in the two platforms (NREL-SAM and HOMER). Accordingly, this project looks into developing models for evaluating power curve tables for Darrieus turbines and synthesizing wind speed scenarios. \\
Analytical and numerical models for Darrieus turbines have been presented, and the ability to evaluate power curve tables from both models has been demonstrated. Double-Multiple Streamtube model was implemented for the analytical modelling, and 2-D CFD simulations on OpenFOAM were developed for the numerical modelling.\\
The analytical model was found to fail for small-sized turbines, likely due to poor lift and drag coefficient values at high local angles of attack, but matched sufficiently with experimental results for large-sized turbines.\\
Numerical simulations for Darrieus turbines were of two types - torque evaluation at constant wind speed and angular velocity, and starting characteristics for constant wind speed.\\
Auto-regressive moving average models were implemented to characterize historical wind speed measurements, thus synthesizing wind speed scenarios. For this, the wind speeds or wind speed residuals (following detrending) were first normalized using the cumulative distribution functions, following which an ARMA model was fit.

\pagebreak

%%%%%%%%%%%%%%%%%%%%%%%%%%%%%%%%%%%%%%%%%%%%%%%%%%%%%%%%%%%%%%%%%
% Table of contents etc.

\begin{singlespace}
\tableofcontents
\thispagestyle{empty}

\listoffigures
\end{singlespace}

%%%%%%%%%%%%%%%%%%%%%%%%%%%%%%%%%%%%%%%%%%%%%%%%%%%%%%%%%%%%%%%%%%%%%%
% Abbreviations
\abbreviations

\noindent 
\begin{tabbing}
xxxxxxxxxxx \= xxxxxxxxxxxxxxxxxxxxxxxxxxxxxxxxxxxxxxxxxxxxxxxx \kill
\textbf{AMI} \> Arbitrary Mesh Interface \\
\textbf{AR} \> Auto-regressive \\
\textbf{ARMA} \> Auto-regressive Moving Average \\
\textbf{BEM} \> Blade Element/Momentum \\
\textbf{CDF} \> Cumulative Distribution Function \\
\textbf{CFD} \> Computational Fluid Dynamics \\
\textbf{DMST} \> Double-Multiple Stream-tube \\
\textbf{HAWT} \> Horizontal Axis Wind Turbine \\
\textbf{HOMER} \> Hybrid Optimization of Multiple Energy Resources \\
\textbf{LA} \> Los Angeles \\
\textbf{MA} \> Moving Average \\
\textbf{NACA} \> National Advisory Committee for Aeronautics \\
\textbf{NOAA} \> National Oceanic and Atmospheric Administration \\
\textbf{NREL} \> National Renewable Energy Laboratory \\
\textbf{PIMPLE} \> PISO-SIMPLE \\
\textbf{PISO} \> Pressure-Implicit with Splitting-Operations \\
\textbf{SAM} \> System Advisor Model \\
\textbf{SD} \> San Diego \\
\textbf{SIMPLE} \> Semi-Implicit Method for Pressure Linked Equations \\
\textbf{SST} \> Shear Stress Transport \\
\textbf{TMY} \> Typical Meteorological Year \\
\textbf{VAWT} \> Vertical Axis Wind Turbine \\
\end{tabbing}

\pagebreak

%%%%%%%%%%%%%%%%%%%%%%%%%%%%%%%%%%%%%%%%%%%%%%%%%%%%%%%%%%%%%%%%%%%%%%
% Notation

\chapter*{\centerline{NOTATION}}
\addcontentsline{toc}{chapter}{NOTATION}

\begin{tabbing}
xxxxxxxxxxx \= xxxxxxxxxxxxxxxxxxxxxxxxxxxxxxxxxxxxxxxxxxxxxxxx \kill
\textbf{$A_k, B_k$} \> Fourier Series Coefficients \\
\textbf{$c$} \> Blade Length, m \\
\textbf{$C_D$} \> Blade Drag Coefficient \\
\textbf{$C_L$} \> Blade Lift Coefficient \\
\textbf{$C_N$} \> Normal Force Coefficient \\
\textbf{$C_P$} \> Turbine Power Coefficient \\
\textbf{$C_T$} \> Tangential Force Coefficient \\
\textbf{$C_{\tau}$} \> Turbine Torque Coefficient \\
\textbf{$H$} \> Turbine Height \\
\textbf{$I_{xx}$} \> Moment of Inertia about x-axis \\
\textbf{$k$} \> Turbulent Kinetic Energy \\
\textbf{$N$} \> Number of Turbine Blades \\
\textbf{$(p,q)$} \> ARMA Model Order \\
\textbf{$R$} \> Turbine Radius, m \\
\textbf{$Re$} \> Reynolds Number \\
\textbf{$S$} \> Turbine Swept Area $=2RH$, m\textsuperscript{2} \\
\textbf{$u_1$} \> Upstream Interference Factor \\
\textbf{$u_2$} \> Downstream Interference Factor \\
\textbf{$V_{d}$} \> Induced Downstream Velocity, m/s \\
\textbf{$V_{e}$} \> Induced Equilibrium (Centre) Velocity, m/s \\
\textbf{$V_{u}$} \> Induced Upstream Velocity, m/s \\
\textbf{$V_{w}$} \> Wake Velocity, m/s \\
\textbf{$V_{\infty}$} \> Free-stream Velocity, m/s \\
\textbf{$W_{u/d}$} \> Upstream/Downstream Local Relative Velocity, m/s \\
\textbf{$W_{sp}$} \> Wind Speed m/s \\
\textbf{$W_F$} \> Wind Speed Trend from Fourier Series \\
\textbf{$W_r$} \> Wind Speed Residual \\
\textbf{$Z$} \> Normalized Wind Speed \\
\\
\textbf{$\alpha_0$} \> Blade Angle of Attack, $^{\circ}$\\
\textbf{$\alpha_{u/d}$} \> Upstream/Downstream Local Angle of Attack, $^{\circ}$ \\
\textbf{$\epsilon$} \> Ch. \ref{chap:cfd}: Turbulent Dissipation \\
                    \> Ch. \ref{chap:arma}: White Noise \\
\textbf{$\theta$} \> Ch. \ref{chap:dmsm}: Blade Azimuth Angle, $^{\circ}$ \\
                  \> Ch. \ref{chap:arma}: MA Parameter \\
\textbf{$\lambda$} \> Tip Speed Ratio $=\omega R/V$ \\
\textbf{$\rho$} \> Density, kg/m\textsuperscript{3} \\
\textbf{$\sigma$} \> Ch. \ref{chap:dmsm} \& \ref{chap:cfd}: Turbine Solidity $=Nc/(2R)$ \\
                  \> Ch. \ref{chap:arma}: Standard Deviation \\
\textbf{$\sigma_{corr}$} \> Correction/Scaling Factor \\
\textbf{$\tau$} \> Torque, N-m \\
\textbf{$\phi$} \> AR Parameter \\
\textbf{$\omega$} \> Ch. \ref{chap:dmsm}: Angular Velocity, rad/s or rpm \\
                  \> Ch. \ref{chap:dmsm}: Specific Turbulence Dissipation Rate \\
\textbf{$\Omega$} \> Ch. \ref{chap:cfd}: Turbine Angular Speed, rad/s \\
                  \> Ch. \ref{chap:arma}: Cumulative Distribution Function\\
\end{tabbing}

\pagebreak
\clearpage

% The main text will follow from this point so set the page numbering
% to arabic from here on.
\pagenumbering{arabic}

%%%%%%%%%%%%%%%%%%%%%%%%%%%%%%%%%%%%%%%%%%%%%%%%%%
% Introduction.

\chapter{INTRODUCTION}
\label{chap:intro}

With an increasing popularization of renewable energy, more and more residential and commercial complexes are looking into developing microgrids and smart grids. As a result, multiple techno-economic models and platforms have been developed to facilitate the decision-making process behind the integration of these renewables. These include the likes of NREL's System Advisor Model (SAM) [\citenum{nrel-sam}], URBANopt and Renewable Energy Integration and Optimization (REopt), and Hybrid Optimization of Multiple Energy Resources (HOMER) [\citenum{homer}]. NREL SAM and HOMER are popular among such platforms in academia, with SAM having over 130,000 users and applications including grid integration studies, policy and utility rate design, and estimate Levelized Cost of Energy for systems. However, while SAM is open-source, HOMER is a paid software and has many subtle additional benefits. In this project, models to complement these platforms have been explored, particularly in wind power performance estimations.

\begin{figure}[htpb]
  \begin{center}
    \subfloat[Screenshot from NREL SAM]{\includegraphics[scale = 0.31]{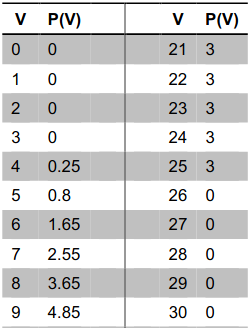}}
    \hfill
    \subfloat[Screenshot from HOMER]{\includegraphics[scale = 0.5]{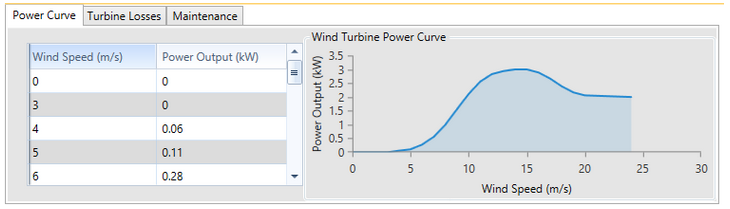}}
    \caption {Power curve tables for wind turbines [\citenum{nrel-sam-wind}, \citenum{homer}].}
  \label{fig:power-sam-homer}
  \end{center}
\end{figure}

SAM and HOMER have models to predict the hourly electrical outputs from wind farms or stand-alone wind turbines using the location's weather data and the turbine's power curve and wake model. For a turbine's power curve, the platforms have libraries of common horizontal axis wind turbines, which contain pre-defined tables of wind speed vs turbine output. Uploading user-defined power curves is also possible by editing these tables. Figure \ref{fig:power-sam-homer} shows screenshots of the power curve tables in NREL SAM and HOMER. 

All the pre-defined tables available in the library are specifically for horizontal axis wind turbines (HAWT), and the alternate vertical axis wind turbines (VAWT) are not included in the library. This is due to the sidelining of VAWT's by the wind energy industry over the years because of the superior efficiencies offered by HAWT's. However, factors including ease of manufacturing, simplicity of design, and the omnidirectional operation of the turbines have led to smaller sized VAWT's gaining popularity, specifically for residential use cases (like rooftop applications). But with limited commercial VAWTs deployed, power curve tables for these turbines are limited. With the intention of accommodating VAWT integrated systems on these platforms, generating power curves using analytical and numerical models have been explored in Chapter \ref{chap:dmsm} and \ref{chap:cfd}, respectively.

The most common design of VAWT's are Darrieus (lift-based) and Savonius (drag-based) turbines, and hybrid turbines (combination of the two) \citenum{dhiman-vawt}. Darrieus turbines consist of multiple aerofoil blades attached to a shaft, and as fluid flows past it, a lift is generated, thus producing torque. These turbines are known to achieve higher power coefficients than other designs, and have thus been chosen for modelling in this project. However, Darrieus turbines are also known to stall at low speeds leading to poor starting torques [\citenum{dhiman-vawt}, \citenum{kirke-phd}]. Thus, the build-up of angular velocity starting from a stationary state has also been modelled in Chapter \ref{chap:cfd}.

Abundant weather data is also a necessary component for evaluating the wind resource potential for a system. Both platforms (SAM \& HOMER) require a weather text file that contains at least one year's worth of data in an hourly or sub-hourly time step for their evaluation and optimization processes. Typical Meteorological Year (TMY), a database maintained by NSRDB that contains yearly data for many locations, is the default weather data loaded on both systems. Besides that, obtaining hourly weather data for a location, particularly for multiple years, is typically complex. HOMER has an additional feature of generating synthetic wind speed data using specific user-defined parameters. However, it does not offer the user features to determine these parameters using available historical data. In Chapter \ref{chap:arma}, models have been developed for generating hourly wind speeds for a location using historical wind speed data for the same location or a location with a similar climate.

\begin{figure}[htpb]
  \begin{center}
    \includegraphics[scale = 0.44]{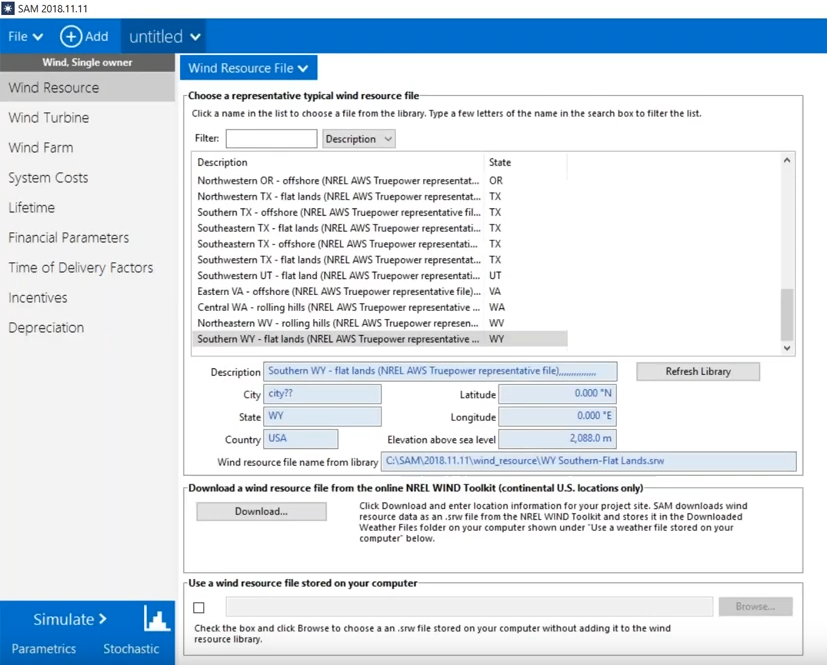}
    \caption {Wind data selection in NREL's SAM [\citenum{nrel-sam-wind}].}
  \label{fig:winddata-sam}
  \end{center}
\end{figure}

\begin{figure}[htpb]
  \begin{center}
    \includegraphics[scale = 0.36]{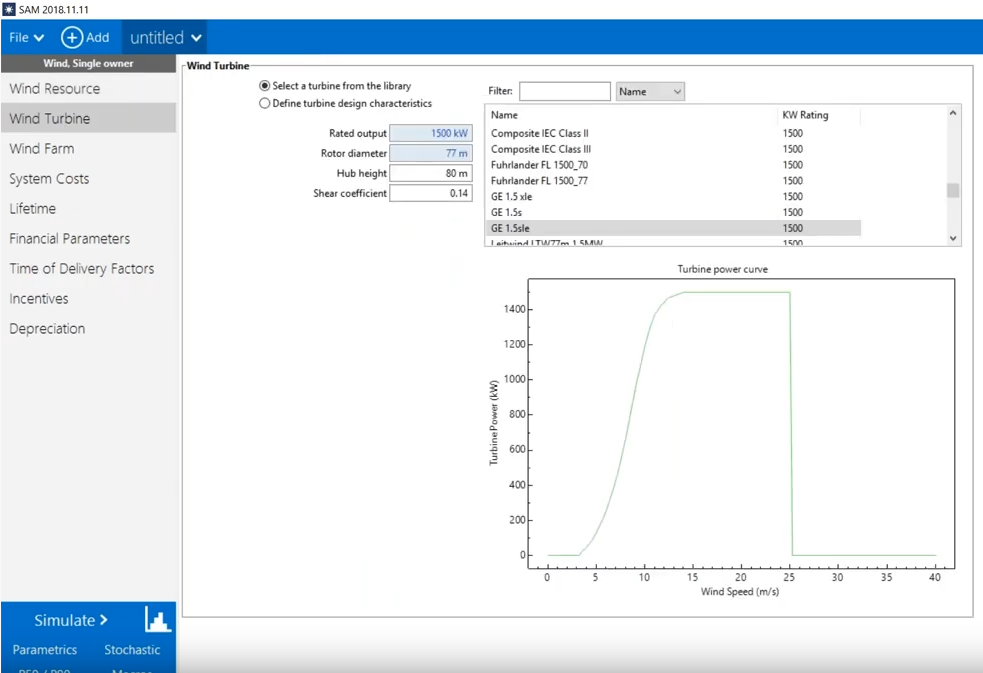}
    \caption {Turbine specification selection in NREL's SAM [\citenum{nrel-sam-wind}].}
  \label{fig:turbinedata-sam}
  \end{center}
\end{figure}

%%%%%%%%%%%%%%%%%%%%%%%%%%%%%%%%%%%%%%%%%%%%%%%%%%
% Analytical Methods.

\chapter{ANALYTICAL MODEL FOR DARRIEUS WIND TURBINES}
\label{chap:dmsm}

Analytical models to characterize vertical axis wind turbines, including both Savonius and Darrieus designs, have been developed since the 1970's. While Darrieus wind turbines can be ``Eggbeater / curved-bladed" or ``Straight-bladed", the models for both are very similar with minor modifications to account for the varying radius and angle of attack across height for the curved-blade geometries. The best validated Darrieus turbine models in literature [\citenum{islam-vawt}] can be classified as Momentum, Vortex and Cascade models.

Momentum or BEM (Blade Element/Momentum) models is based on calculating the flow velocity through the turbine by equating the aerodynamic force with the change in momentum of air flow. These models are however known to fail at large values of solidity, $\sigma$.

The first and simplest BEM model developed was the single streamtube model by \cite{templin}. In this, the entire turbine is assumed to be enclosed within a single streamtube, and a uniform flow velocity within the swept area equal to the average of the free-stream and wake velocity is considered. This model is applicable for only lightly loaded turbines, and often over-predicts the produced torque.

The multiple streamtube model was first proposed by \cite{wilson-vawt} where the swept area was divided into a series of adjacent and aerodynamically independent streamtubes. It was further developed by \cite{strickland-vawt} incorporating the wind shear effects.

The double-multiple streamtube model was proposed by \cite{dmsm-vawt}, where the turbine rotation was divided into upstream and downstream half-cycles, thus ``doubling" the number of streamtubes. This is the model chosen for the project, and is explained in the following section.

Vortex models are potential flow models based on calculating the velocity field about the turbine through the influence of vorticity in blade wakes. It was first proposed for 2-D analysis of VAWT's by \cite{vortex-vawt}. The turbine blades are represented by bound or lifting-line vortices whose strengths are determined using the airfoil coefficient for the calculated relative flow velocity and angle
of attack. These models are accurate for only small local angles of attack, and are also known to take the most computational time.

Cascade models are commonly used for other turbo-machines, and was first applied to Darrieus turbines by \cite{cascade-vawt}. In the model, the blades are rolled-out onto a cascade (a plane normal to the turbine axis) with the interspace equal to the circumferential distance between two adjacent blades. After calculating the local angles of attack and flow relative velocity, the blades are arranged in the cascade and the aerodynamic forces are evaluated for a revolution of the reference blade. This model is known to arrive at solutions much faster than the vortex model and is applicable for a wider range of solidity than the BEM models.

\section{Double-Multiple Streamtube Model}

An analytical model for Darrieus wind turbine has been developed in Python following \cite{dmsm-vawt}. While [\citenum{dmsm-vawt}] was developed keeping curved-blade turbines in mind, the model has been implemented for straight-blade turbines in this report. [\citenum{dmsm-vawt}] also considers tall turbines with local free-stream velocities varying as a function of height - this has been ignored in this project.

\subsubsection*{Induced velocities and local relative velocities}

\begin{figure}[htpb]
  \begin{center}
    \includegraphics[scale = 0.43]{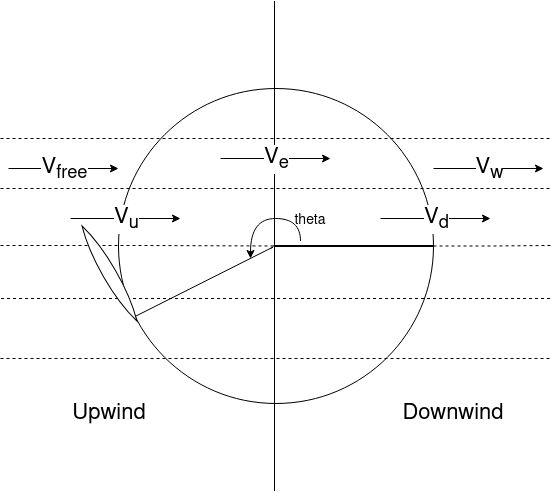}
  \caption {Representation of Double-Multiple Streamtube model for a single blade}
  \label{fig:aero-dmsm}
  \end{center}
\end{figure}

A 2-D sketch of airflow past a blade of a Darrieus turbine has been presented in Fig. \ref{fig:aero-dmsm}. As presented, the flow past the turbine has been divided into multiple ``streamtubes", each having an upstream and a downstream section. Along the streamtube, different induced velocities have been identified ($V_{u}$, $V_{e}$ and $V_{d}$) assuming two actuator discs in tandem [\citenum{act-discs}]. 

$V_{u}$ and $V_{d}$ are averages of $V_\infty$ (free-stream velocity) and $V_{e}$, and $V_{e}$ and $V_{w}$ (wake velocity) respectively. Interference factors to account for the decrease in velocities at the upstream and downstream discs have been introduced - $u_1$ and $u_2$, and their values are determined in an iterative manner as described later ahead. Thus the induced velocities become:
\begin{equation}
    V_u = u_1 V_{\infty}
\end{equation}
\begin{equation}
    V_e = 2 V_u - V_{\infty} = (2 u_1 - 1) V_{\infty}
\end{equation}
\begin{equation}
    V_d = u_2 V_e = u_2 (2 u_1 - 1) V_{\infty}
\end{equation}

The upstream half-cycle of the rotor is for Azimuth angles $-\pi/2 < \theta < \pi/2$. In this region, the local relative velocity at the blade tip is given by:
\begin{equation}
    W_u = V_u\sqrt{(\lambda_u - \sin{\theta})^2 + \cos^2{\theta}}
\label{eqn:locWs}
\end{equation}
and the local angle of attack is given by:
\begin{equation}
    \alpha_u = \arcsin{\left[ \dfrac{\cos{\theta} \cos{\alpha_0} - (\lambda_u - \sin{\theta}) \sin{\alpha_0}}{\sqrt{(\lambda_u - \sin{\theta})^2 + \cos^2{\theta}}} \right]}
\label{eqn:locAoA}
\end{equation}
where $\alpha_0$ is the blade's angle of attack w.r.t. the tangential direction. $\lambda_u$ is the upstream tip speed ratio given by:
\begin{equation}
    \lambda_u = \omega R / V_u
\label{eqn:locTSR}
\end{equation}

The downstream half-cycle of the rotor is for Azimuth angles $\pi/2 < \theta < 3\pi/2$. The local angle of attack ($\alpha_d$), relative flow velocity ($W_d$) and tip speed ratio ($\lambda_d$) can be determined from Equations \ref{eqn:locWs}, \ref{eqn:locAoA} and \ref{eqn:locTSR} using $V_d$ instead of $V_u$.

\subsubsection*{Force Coefficients}

Due to the flow past the blades, an elemental lift and drag force is generated. The tangential and normal coefficients for a blade is given by:
\begin{equation}
    C_N = C_L \cos{\alpha} + C_D \sin{\alpha}
\end{equation}
\begin{equation}
    C_T = C_L \sin{\alpha} - C_D \cos{\alpha}
\end{equation}
$C_L$ and $C_D$ are the blade's lift and drag coefficients, and is a function of both - the local Reynolds Number ($Re$) and the local angle of attack ($\alpha$). The local $Re$ for a given streamtube can be evaluated as:
\begin{equation}
    Re = \dfrac{V_{u/d} c}{\nu_{\infty}} \sqrt{(\lambda_{u/d} - \sin{\theta})^2 + \cos^2{\theta}}
\end{equation}
The values for $C_L$ and $C_D$ have been interpolated from the values in [\citenum{osti-clcd}]. Other sources for these values include Xfoil [\citenum{xfoil-program}] and \url{airfoiltools.com}. 

\subsubsection*{Interference factor evaluation}

As stated, the interference factors, $u_1$ and $u_2$ are iteratively evaluated. The convergence criteria from BEM theory for them are:
\begin{equation}
    \dfrac{\pi(1 - u_1)}{u_1} = \dfrac{N c}{8 \pi R} \int_{-\pi/2}^{\pi/2} \left( C_{N,u} \dfrac{\cos{\theta}}{|\cos{\theta}|} - C_{T,u} \dfrac{\sin{\theta}}{|\cos{\theta}|} \right) \left( \dfrac{W_u}{V_u} \right)^2 d\theta
\label{eqn:conv-up}
\end{equation}
\begin{equation}
    \dfrac{\pi(1 - u_2)}{u_2} = \dfrac{N c}{8 \pi R} \int_{\pi/2}^{3\pi/2} \left( C_{N,d} \dfrac{\cos{\theta}}{|\cos{\theta}|} - C_{T,d} \dfrac{\sin{\theta}}{|\cos{\theta}|} \right) \left( \dfrac{W_d}{V_d} \right)^2 d\theta
\label{eqn:conv-dw}
\end{equation}
It is taken that $0 < u_2 < u_1 < 1$, and their values decrease with increased tip speed ratio. By starting with an assumption of $u_1 = 1$, $u_1$ can be iteratively determined. Following it, starting with $u_2 = u_1$, $u_2$ can also be determined iteratively.

\subsubsection*{Torque and power coefficient evaluation}

The average torque produces in one revolution is determined as:
\begin{equation}
    \Bar{\tau} = \dfrac{N}{2\pi} \left[ \int_{-\pi/2}^{\pi/2} \tau_u(\theta) d\theta + \int_{\pi/2}^{3\pi/2} \tau_d(\theta) d\theta \right]
\end{equation}
where,
\begin{equation}
    \tau_u (\theta) = \frac{1}{2} \rho_{\infty} c R H C_{T,u} W_u^2
\end{equation}
\begin{equation}
    \tau_d (\theta) = \frac{1}{2} \rho_{\infty} c R H C_{T,d} W_d^2
\end{equation}

From the above, the over-all power coefficient for the turbine can be determined as:
\begin{equation}
    C_P = \dfrac{N c \lambda H}{2 \pi S} \left[ \int_{-\pi/2}^{\pi/2} C_{T,u} \left(\dfrac{W_u}{V_{\infty}}\right)^2 d\theta + \int_{\pi/2}^{3\pi/2} C_{T,d} \left(\dfrac{W_d}{V_{\infty}}\right)^2 d\theta \right]
\end{equation}
where, $S$ is the swept area - which for a straight-bladed turbine is $2 R H$.

\section{Results and Discussions}

\subsubsection*{Results for large turbines}

Using the above procedure, double-multiple streamtube (DMST) model for a straight-bladed Darrieus turbine has been developed in Python. The turbine is considered to be of $R = 2.5$ m and $\sigma = 0.22$ with 3 NACA0015 blades, similar to SANDIA's 5-m curved-bladed turbine [\citenum{sandia-vawt}]. Experimental data for this turbine is available for $\omega = 150$ rpm. Thus, for comparison purposes the results from the DMST model for the same angular speed have been considered. 

\begin{figure}[htpb]
  \begin{center}
    \subfloat[Tangential force coefficient]{\includegraphics[scale = 0.7]{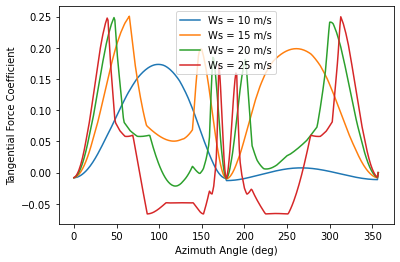}}
    \\
    \subfloat[Individual blade torque]{\includegraphics[scale = 0.7]{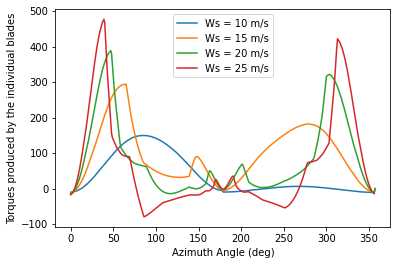}}
    \\
    \subfloat[Total turbine torque for $V_{\infty} = 15$m/s]{\includegraphics[scale = 0.7]{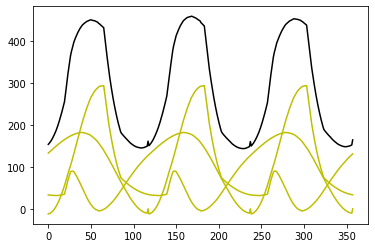}}
    \caption {Variation in local $C_T$ and $\tau$ for varying $\theta$.}
  \label{fig:anlyt-theta-var}
  \end{center}
\end{figure}

Figure \ref{fig:anlyt-theta-var} (a) and (b) present the variation in the tangential force coefficient ($C_T$) and torque produced by individual blade ($\tau_i$) as a function of Azimuth angle ($\theta$). A negative torque being produced at certain intervals of $\theta$ can be observed in the graph. However, due to the influence of other blades with positive torques, a positive net power is produced by the turbine. This can be seen in Fig. \ref{fig:anlyt-theta-var} (c).

\begin{figure}[htpb]
  \begin{center}
    \subfloat[Experimental $C_P$-$\lambda$ data]{\includegraphics[scale = 0.3]{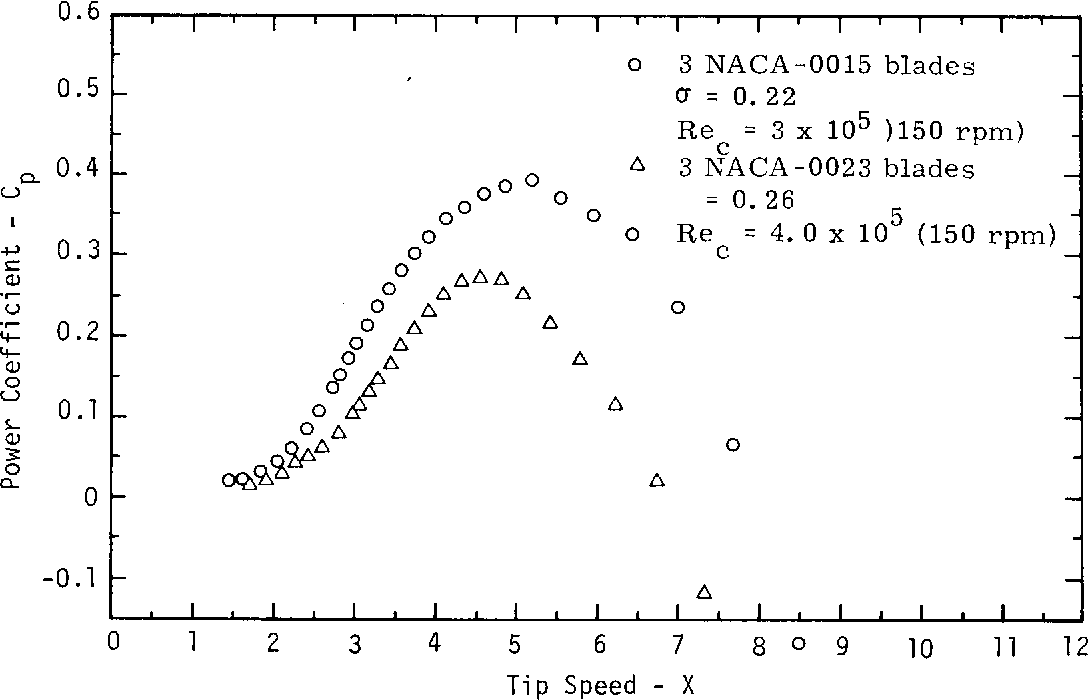}}
    \\
    \subfloat[Analytical $C_P$-$\lambda$ solution]{\includegraphics[scale = 0.7]{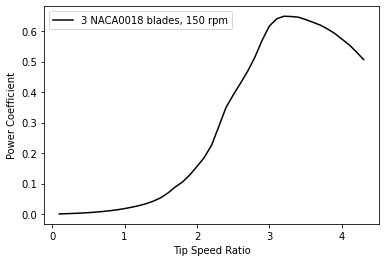}}
    \\
    \subfloat[Analytical $C_{\tau}$-$\lambda$ solution]{\includegraphics[scale = 0.7]{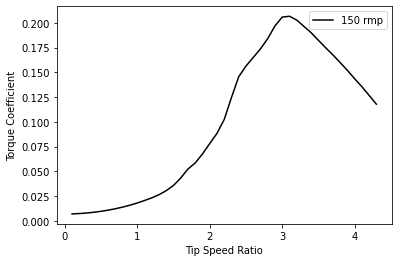}}
    \caption {Comparison with experimental data for SANDIA's $5$m curved-bladed VAWT.}
  \label{fig:sandia-comp}
  \end{center}
\end{figure}

Figure \ref{fig:sandia-comp} presents a comparison between the $C_P$ calculated using the DMST model and the experimental $C_P$ obtained from SANDIA's 5-m turbine [\citenum{sandia-vawt}]. It can be seen that the two graphs present similar characteristics. It must be noted that since the experimental data is for a curve-bladed turbine, a higher $C_P$ and a shift in the peak is expected from the analytical solution. We can thus say that the analytical model is able to predict the Darrieus turbine performance with sufficient accuracy for large turbines. Figure \ref{fig:sandia-comp} (c) presents the Torque coefficient, $C_{\tau}$, obtained for the turbine using the DMST model.

\begin{figure}[htpb]
  \begin{center}
    \includegraphics[scale = 0.7]{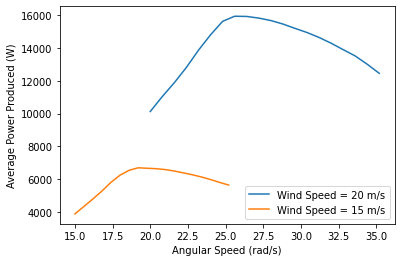}
    \caption {Power produced for varying angular speeds}
  \label{fig:anlyt-P-comp}
  \end{center}
\end{figure}

Fig. \ref{fig:anlyt-P-comp} presents the power variation obtained for the same turbine operating at the wind speed of $V_{\infty} = 15$ and $20$m/s. This allows us (apart from directly evaluating from the $C_P$ curve) to determine the optimal power that can be harvested for a given wind speed - which is what is required for softwares such as HOMER and NREL SAM.

\begin{figure}[htpb]
  \begin{center}
    \includegraphics[scale = 0.7]{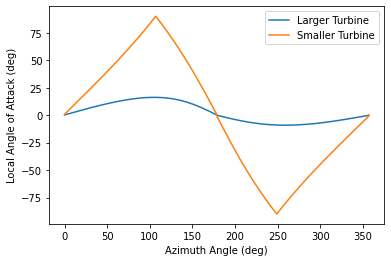}
    \caption {Variation in local angle of attack against turbine size}
  \label{fig:AoA-comp}
  \end{center}
\end{figure}

\begin{figure}[htpb]
  \begin{center}
    \includegraphics[scale = 0.7]{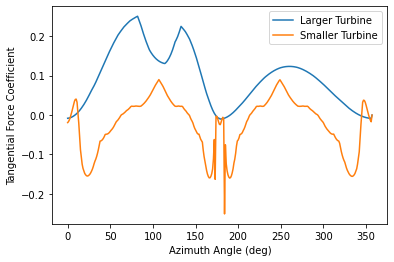}
    \caption {Variation in tangential force coefficient against turbine size}
  \label{fig:CT-comp}
  \end{center}
\end{figure}

When small sized VAWT's are considered, the model was found to fail. A possible reason for this could be the high local angles of attacks encountered by small turbines. As stated, experimental data from [\citenum{osti-clcd}] was used for the lift and drag coefficients in the model. These values were listed for discrete values of Reynolds number and angle of attack, and the model interpolates accordingly. Due to a high variation in $C_{L/D}$ and larger separation between consecutive $\alpha$'s, these interpolations are likely erroneous leading to the model's poor performance. Figure \ref{fig:AoA-comp} presents a comparison between the calculated angles of attack encountered by a blade of two sizes of turbines. Figure \ref{fig:CT-comp} presents a comparison between the calculated tangential force coefficients. The parameter values for the two turbine simulations are:
\begin{itemize}[leftmargin = 8em]
    \item[Larger Turbine:] $R = 2.5$m, $\sigma = 0.22$, $V_{\infty}  = 15$m/s, $\omega = 18$ rad/s
    \item[Smaller Turbine:] $R = 0.3$m, $\sigma = 0.3$, $V_{\infty}  = 10$m/s, $\omega = 10$ rad/s
\end{itemize}

\begin{figure}[htpb]
  \begin{center}
    \subfloat[Variation with Angle of Attack]{\includegraphics[scale = 0.8]{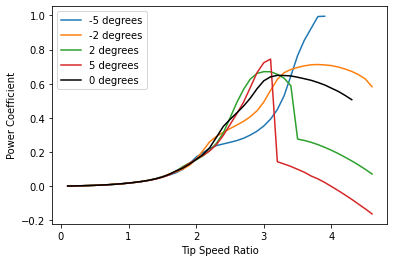}}
    \\
    \subfloat[Variation with solidity]{\includegraphics[scale = 0.8]{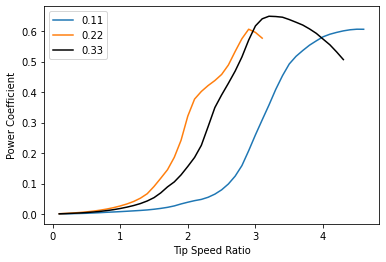}}
    \caption {Variation in $C_P$ for variation in $\alpha_0$ and $\sigma$}
  \label{fig:Cp-anlyt-comp}
  \end{center}
\end{figure}

Since analytical models require much less computational time than CFD models discussed in Chapter \ref{chap:cfd}, these models are beneficial for parametric studies - though not relevant in the context of techno-economic models. Fig. \ref{fig:Cp-anlyt-comp} presents the variation in $C_P$ for varying solidity ($\sigma$) and blade angles of attack ($\alpha_0$) as an example. The plots obtained are similar to those in [\citenum{kirke-phd}]. Continuing, sudden drops in $C_{P}$ are observed in Fig. \ref{fig:Cp-anlyt-comp} (a). This is once again likely due to the erroneous interpolations for $C_L$ and $C_D$, as was observed for small-sized turbines.

%%%%%%%%%%%%%%%%%%%%%%%%%%%%%%%%%%%%%%%%%%%%%%%%%%
% Numerical Methods.

\chapter{NUMERICAL MODELS FOR DARRIEUS WIND TURBINES}
\label{chap:cfd}

For a more detailed analysis to complement the analytical model, numerical simulations have also been performed. The C++ based CFD toolkit ``OpenFOAM" [\citenum{openfoam}] has been adopted for this purpose. Being an open-source package with entirely text-based interface, OpenFOAM is particularly advantageous than some of the other common CFD softwares. 2-D CFD simulations on Darrieus turbines of two kinds have been performed - torque evaluation for a constant tip speed ratio ($\lambda$), and angular speed build up from a stationary state. IITM HPCE's \href{https://hpce.iitm.ac.in/content.php?navigate=aquacluster}{AQUA Cluster} has been used for the simulations.

\section{Overview of the OpenFOAM toolbox}

Open source Field Operation And Manipulation (OpenFOAM) is a cost-free, open–source CFD package developed primarily by the OpenCFD Ltd to resolve flows numerically and provide different solvers, libraries, and utilities for various CFD problems. The OpenFOAM package offers a robust and adaptable simulation platform and incorporates a C++ toolbox for the development of personalized numerical solvers and pre-/post-processing utilities for the solution of continuum mechanics problems.

\subsubsection*{PimpleFoam solver}

OpenFOAM offers a variety of solvers that can be used for the simulations, ranging from electromagnetics, finances and chemical reactions to compressible and incompressible flows. The solver \verb+pimpleFoam+ has been selected for the Darrieus turbine modeling, which deals with transient incompressible flows. The solver uses the PIMPLE algorithm which combines the PISO and SIMPLE algorithms. In the older versions of OpenFOAM (until V6.0), a separate solver - pimpleDyMFoam was required for situations with dynamic meshes, and can still be seen in most of the available examples. 

\begin{figure}[htpb]
  \begin{center}
    \includegraphics[scale = 0.55]{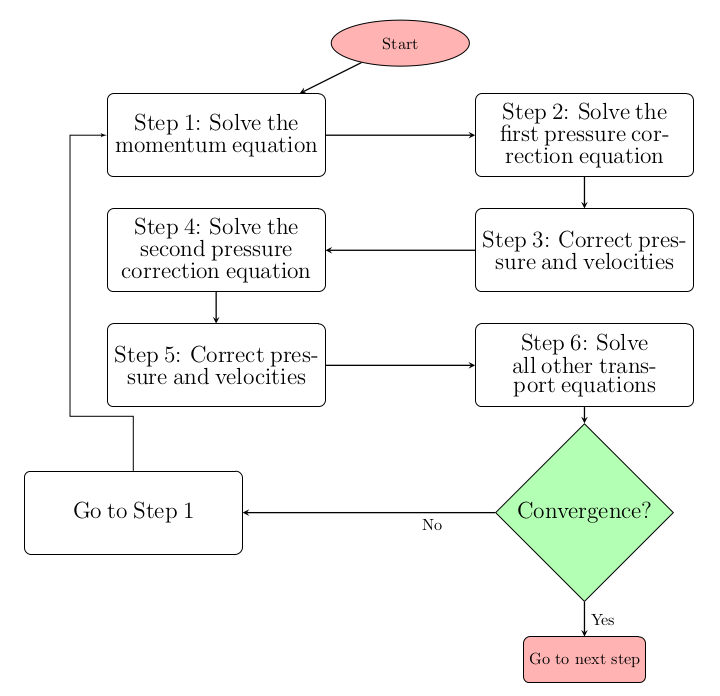}
  \caption {PIMPLE implementation as illustrated in [\citenum{lux-ms}]}
  \label{fig:pimple_fc}
  \end{center}
\end{figure}

SIMPLE (Semi-Implicit Method for Pressure-Linked Equations) [\citenum{simple-caretto}] is an iterative numerical procedure for solving Navier-Stokes equations, and is meant for steady state flows alone. In it, the discretized momentum equation and the pressure correction equation are solved approximate velocity field, then the pressure distribution, and finally the corrected velocity, and the processes is iterated until convergence [\citenum{simple-wiki}]. PISO (Pressure Implicit with Splitting of Operators) [\citenum{piso-issa}] is an extension of the SIMPLE algorithm, and includes a predictor step followed by two corrector step, and is applicable for transient flows too. Implementation of the PIMPLE algorithm has been illustrated in Fig. \ref{fig:pimple_fc}.

\subsubsection*{Turbulence model}

Flow around the turbine is an evidently turbulent flow, thus it is key to model its effects on the system. $k-\epsilon$ [\citenum{turb-k-eps}] and $k-\omega$ [\citenum{turb-k-om}] are common turbulence models used in CFD simulations, and are of the two-equations type. By definition, two equation models include two additional transport equations for representing the turbulent properties of the flow. The two additional transported variables are the turbulent kinetic energy, $k$, and one of turbulent dissipation, $\epsilon$, or the specific turbulence dissipation rate, $\omega$, based on the model. The $k-\epsilon$ model is known to have poor performance in simulations with high pressure gradients or complex flows with strong curvatures. The $k-\omega$ model is the preferred model for low $Re$ flows and best used for near-wall treatment, and is very sensitive to free-stream turbulence. In this project, the $k-\omega$ SST model [\citenum{SST-k-om}], which is known to ``combine best of worlds ($k-\epsilon$ and $k-\omega$)" has been chosen for modeling the turbulence. The SST (shear stress transport) formulation emphasizes on the $k-\omega$ behaviour in the inner parts of the boundary layer and switches to the $k-\epsilon$ behaviour in the free-stream. The drawbacks of the model is over prediction of turbulence at stagnation regions, which are less pronounced for Darrieus turbines, and more computational requirements.

\section{Geometry and Mesh Creation}

The first step of a CFD simulation is the ``pre-processing". This includes creating or deciding a geometry for the simulation, and following it generating a mesh system that can be used by the solver. 

\subsubsection*{Geometry Generation}

\begin{figure}[htpb]
  \begin{center}
    \includegraphics[scale = 0.5]{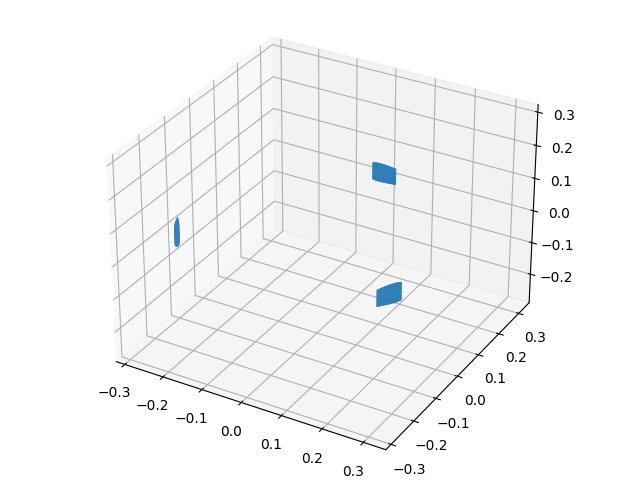}
    \caption {STL files generated using numpy-stl and python.}
  \label{fig:stl_gen}
  \end{center}
\end{figure}

STL is a convenient and typical format used for 3-D geometries. In an STL file, a raw triangulated surface of the geometry is represented with a combination of the normal and vertices of these triangles. Generating the relevant STL files for the Darrieus turbine has been simplified using the Python library \verb+numpy-stl+ [\citenum{numpy-stl}]. The Python file \verb+create_blade+ generates an STL file for a single NACA blade as per the inputs. These inputs include the blade profile, number of vertices, blade height (z-axis), chord length and angle of attack. Following it, the python file \verb+create_VAWT+ uses the generated \verb+blade.stl+ file to generate \verb+geometry.stl+ which represents the Darrieus turbine \footnote{3-D modeling softwares such as Fusion360, FreeCAD and CREO can be tedious for creating multiple geometries with varying blade profiles. In comparison, using Python, the geometry is generated in a matter of seconds.}. The \verb+numpy-stl+ library is also used to evaluate the created geometry's volume and inertia tensor, which proves useful for simulating the turbine's starting characteristics. Figure \ref{fig:stl_gen} presents a generated Darrieus turbine using 3 NACA0018 airfoils. The \verb+geometry.stl+ file is then converted to a 2-D mesh, as described ahead, allowing it to be used by the solver.

\subsubsection*{Mesh Generation}

\begin{figure}[htpb]
  \begin{center}
    \includegraphics[scale = 0.25]{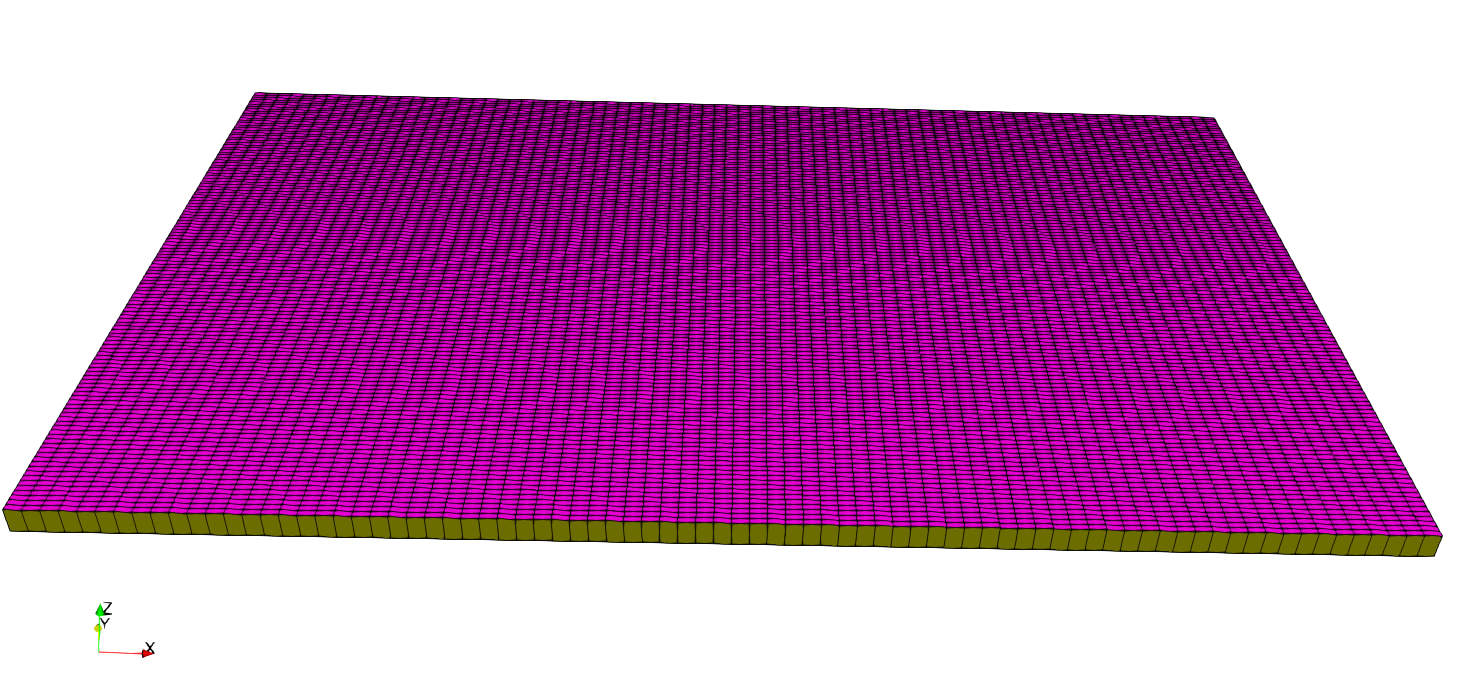}
  \caption {2-D background mesh generated using blockMesh}
  \label{fig:blockMesh}
  \end{center}
\end{figure}

Creating a background mesh is a key component of the pre-processing stage. A simple cuboid of appropriate dimensions (sufficiently larger than the turbine radius) and divisions is created using \verb+blockMeshDict+ for this purpose. The left face is declared as an inlet patch and right the outlet patch, while the rest are simple walls. Since we are proceeding with a 2-D simulation, the z-direction is one cell/division thick, and a thickness of $0.05$m is arbitrarily chosen. Fig. \ref{fig:blockMesh} presents a generated background mesh.

\begin{figure}[htpb]
  \begin{center}
    \subfloat[Single Blade]{\includegraphics[scale = 0.6]{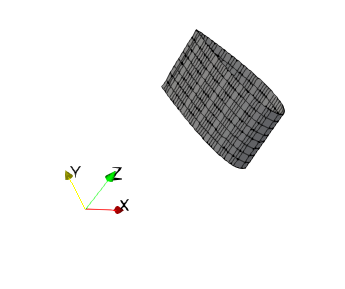}}
    \hfill
    \subfloat[AMI Region]{\includegraphics[scale = 0.3]{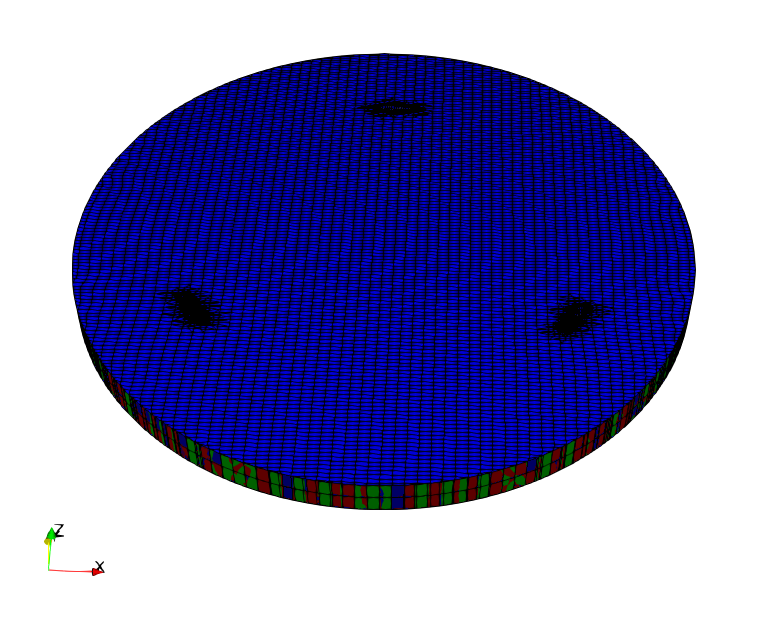}}
    \caption {3D mesh generated from stl using SnappyHexMesh.}
  \label{fig:VAWT_mesh}
  \end{center}
\end{figure}

The STL files are converted into a 3-D mesh using the surfaceFeature and SnappyHexMesh features of OpenFoam. SurfaceFeatures uses the \verb+surfaceFeaturesExtract+ file to extract usable stuff and create \verb+.emesh+ files within the triSurface folder. Following it, SnappyHexMesh creates them into 3-D meshes. There are many constraints and conditions you have to add to it. Fig. \ref{fig:VAWT_mesh} shows an example of SnappyHexMesh applied on an airfoil.

\begin{figure}[htpb]
  \begin{center}
    \subfloat[Single Blade]{\includegraphics[scale = 0.6]{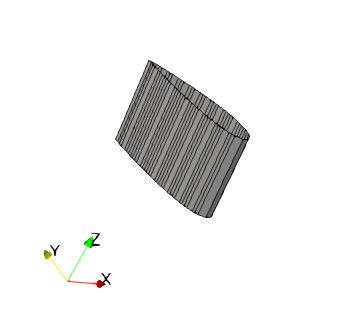}}
    \hfill
    \subfloat[AMI Region]{\includegraphics[scale = 0.3]{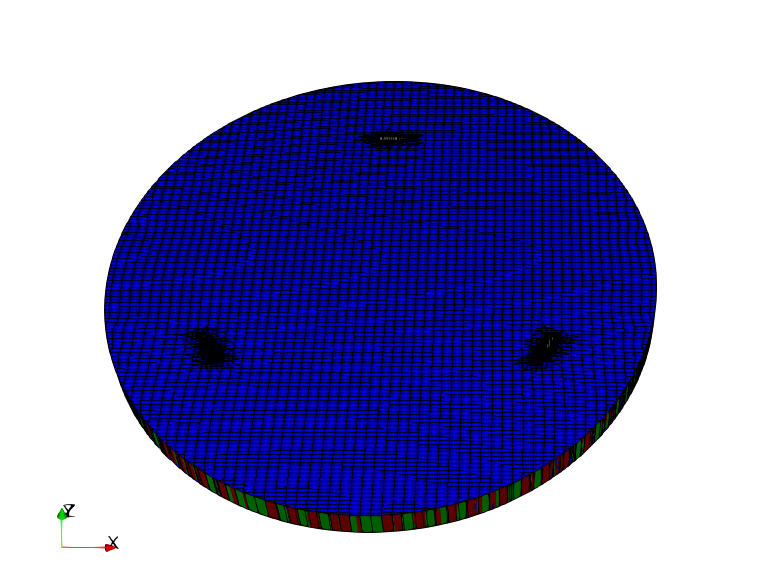}}
    \caption {2D mesh generated from 3D using extrudeMesh.}
  \label{fig:extr_mesh}
  \end{center}
\end{figure}

Since SnappyHexMesh provides a 3-D mesh, it needs to be converted to a 2-D equivalent with a uniform 1-cell thickness throughout the body. For this purpose, the \verb+ExtrudeMesh+ feature is used. ExtrudeMesh selects a single layer of vertices, and extrudes it by a fixed thickness. The thickness would be the same as the background mesh. Fig. \ref{fig:extr_mesh} shows the meshes from Fig. \ref{fig:VAWT_mesh} converted from 3-D to 2-D.

\begin{figure}[htpb]
  \begin{center}
    \subfloat[Top view of mesh at the start]{\includegraphics[scale = 0.4]{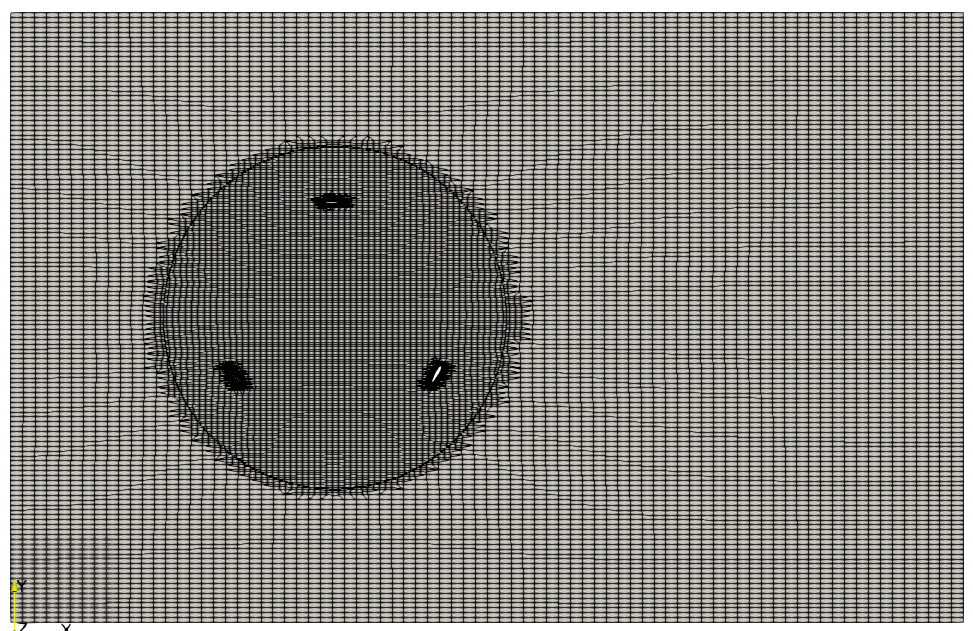}}
    \\
    \subfloat[Flow velocity distribution at an instant]{\includegraphics[scale = 0.3]{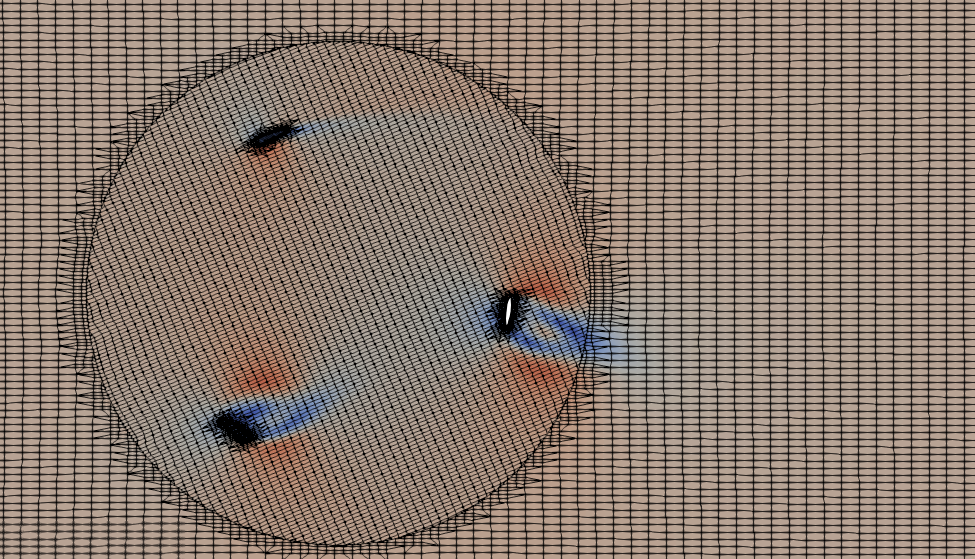}}
    \\
    \subfloat[Flow velocity distribution at an instant]{\includegraphics[scale = 0.3]{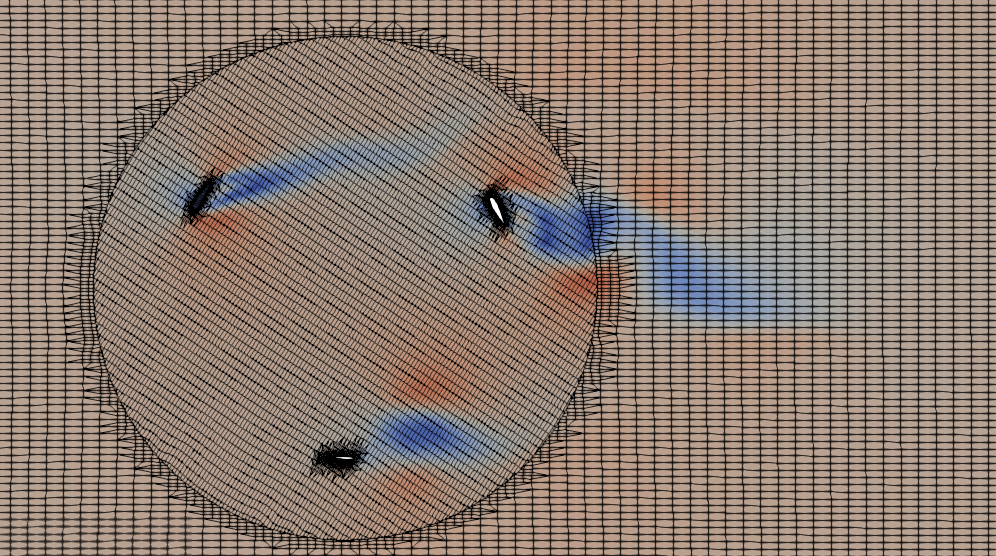}}
    \caption {Final 2-D mesh for the system}
  \label{fig:full_mesh}
  \end{center}
\end{figure}

Since the simulation has a moving component (the turbine), a dynamic mesh is required. For this purpose, the arbitrary mesh interface (AMI) feature of OpenFOAM is used along with a dynamicMeshDict file. The circular region in Fig \ref{fig:extr_mesh} is the AMI region, and is generated using an \verb+AMI.stl+ file through a process similar to \verb+geometry.stl+. The size of the AMI region is chosen as roughly equidistant from the turbine geometry and the background mesh's boundary. The rate at which the AMI region, thus the turbine, rotates is determined by the dynamicMeshDict file. For the constant $\lambda$ simulation, a constant angular speed is assigned, whereas for starting characteristics, the angular speed is determined using the 6-dof library. The top-view of the final mesh as displayed on paraView has been presented in Fig. \ref{fig:full_mesh}. The left face of the outer cuboid is the inlet patch for the system, and the right face is the system's outlet patch. Following the mesh generation, the pimpleFoam solver is called and the system is solved for each instant. A closer look at the AMI-background interface shows that there are no free-edges, allowing the pimpleFoam solver to converge at each time instant across the entire mesh.

\section{Evaluation of Torque at constant ($W_{sp}$, $\Omega$)}

In order to evaluate the $C_p$ and power curve for the given Darrieus turbine, it is important to evaluate the torques at various values of wind speeds ($W_{sp}$) and angular speeds ($\Omega$). An OpenFOAM case for this purpose has been included in the \verb+CO_VAWT+ folder. A VAWT of radius $0.3$ m with solidity $0.3$ and 3 NACA0018 blades has been selected for the following simulations. In order to modify it for different situations, the following can be done:
\begin{itemize}
    \item To change the geometry of the VAWT, the \verb+geometry.stl+ file in \verb+constant/triSurface/+ must be replaced.
    \item Should the new turbine geometry be too big or small, the background mesh and AMI region too must be modified. This can be done by modifying the \verb+blockMeshDict+ file in \verb+system/+ and resizing \verb+AMI.stl+ in \verb+constant/triSurface/+ respectively.
    \item \verb+omega+ in \verb+constant/dynamicMeshDict+ file must be changed.
    \item \verb+0.orig/U+ must be modified with the appropriate wind speed.
\end{itemize}

\begin{figure}[htpb]
  \begin{center}
    \includegraphics[scale = 0.43]{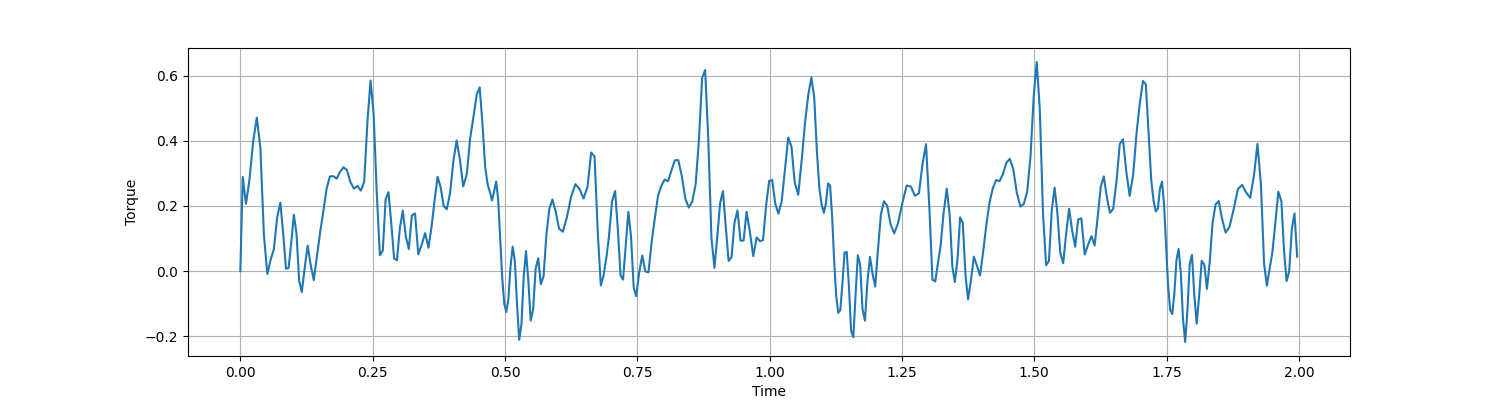}
  \caption {Torque generated for ($W_{sp} = 10$ m/s, $\Omega = 10$ rad/s)}
  \label{fig:torque-eg}
  \end{center}
\end{figure}

Using the \verb+forces+ library, the forces and torque experienced by the VAWT mesh is tabulated for each time-step. The forces and moments due to pressure, viscous and porous effects across the 3 directions is tabulated in \verb+forces.dat+. The sum of moments due to pressure and viscous effects about the z-axis gives us the torque required for our calculations. It must be noted however, that these torques need to be multiplied by $20H$ to find the actual torque, as the simulations are for blade height of $0.05$ m. The instantaneous torque values have been extracted into \verb+torque.csv+ and then plotted using python. Fig. \ref{fig:torque-eg} presents the torque variation obtained for ($W_{sp} = 10$ m/s, $\Omega = 10$ rad/s) taking $H = 1$ m.

\begin{figure}[htpb]
  \begin{center}
    \subfloat[$\Omega = 5$ rad/s]{\includegraphics[scale = 0.43]{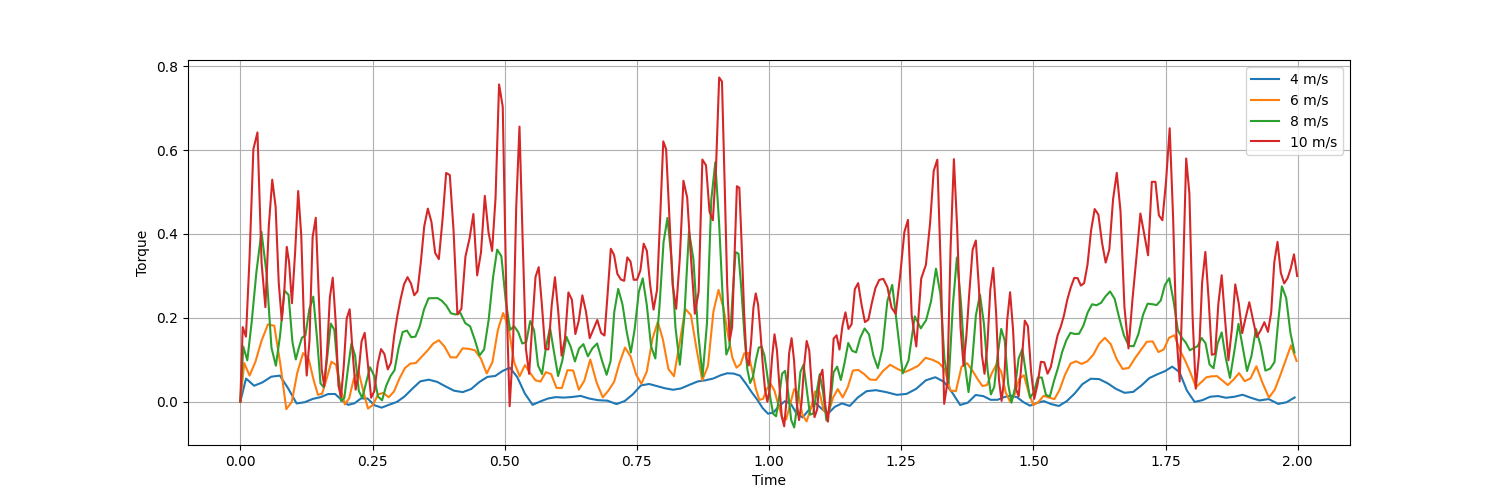}}\\
    \subfloat[$\Omega = 7.5$ rad/s]{\includegraphics[scale = 0.43]{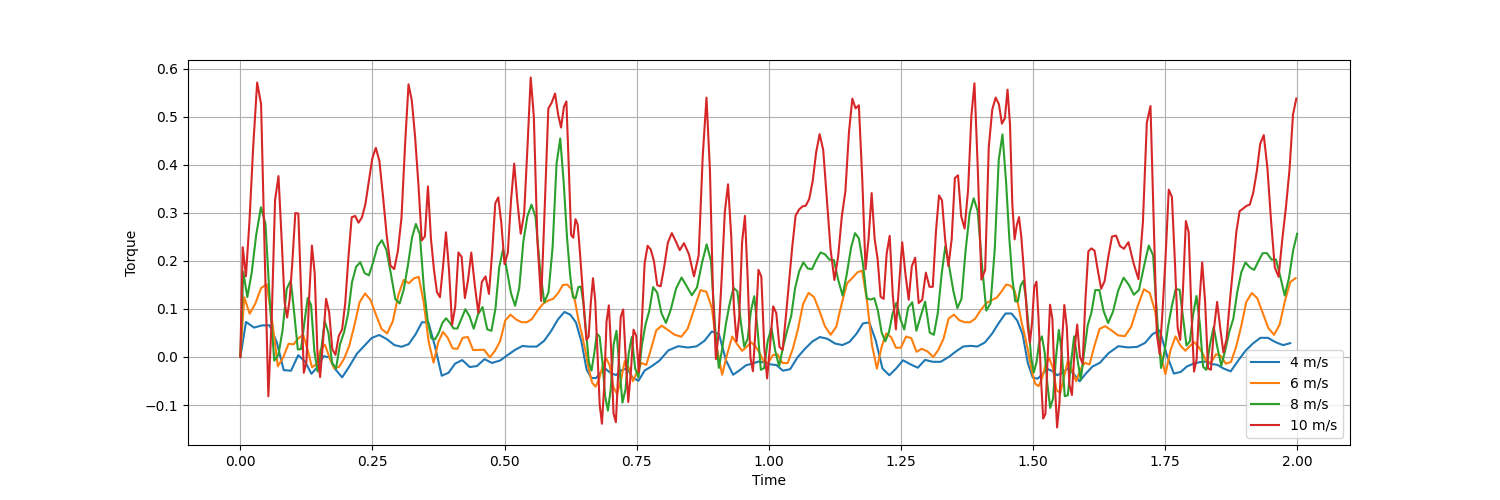}}\\
    \subfloat[$\Omega = 10$ rad/s]{\includegraphics[scale = 0.43]{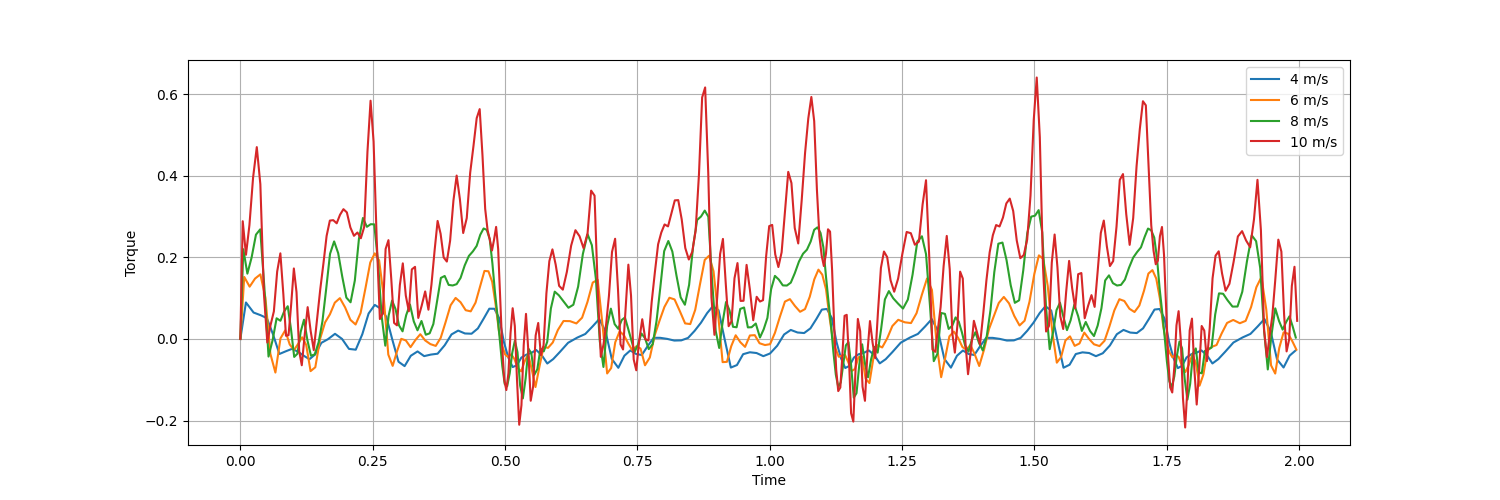}}
    \caption {Comparison of the instantaneous torques for varied ($W_{sp},\Omega$)}
  \label{fig:torque-comp}
  \end{center}
\end{figure}

\begin{figure}[htpb]
  \begin{center}
    \subfloat[Power produced for varying ($W_{sp}, \Omega$)]{\includegraphics[scale = 0.55]{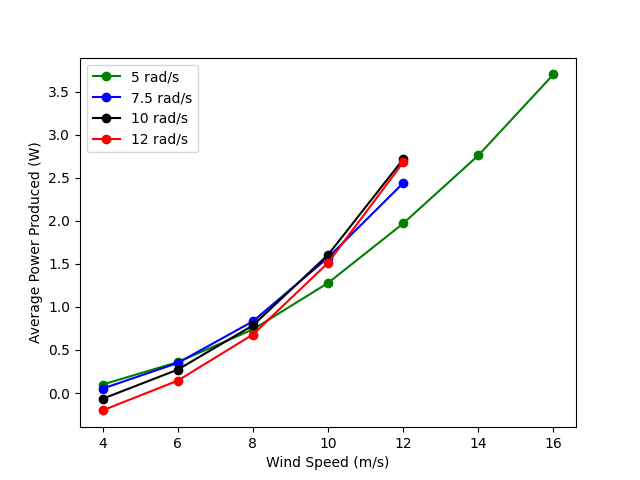}}\\
    \subfloat[Torque produced for varying ($W_{sp}, \Omega$)]{\includegraphics[scale = 0.55]{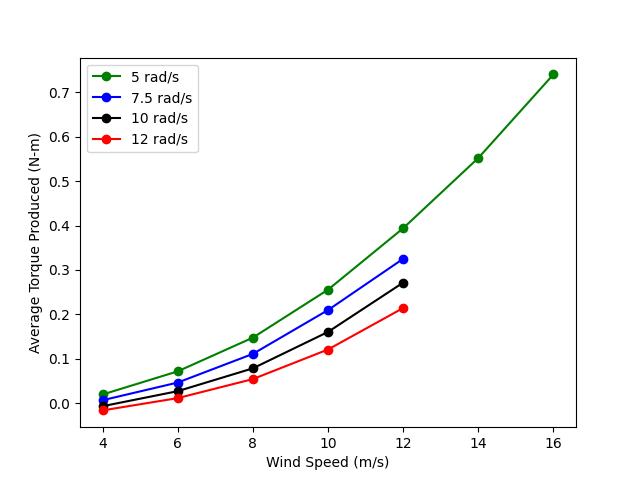}}\\
    \subfloat[Power produced for $W_{sp}$ fixed at 8 and 10 m/s]{\includegraphics[scale = 0.66]{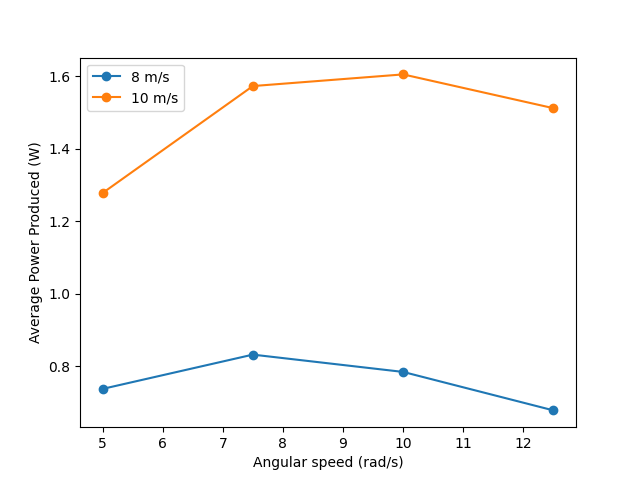}}
    \caption {Comparison of average torques and powers produced}
  \label{fig:power-comp}
  \end{center}
\end{figure}

Figure \ref{fig:torque-comp} presents the torque generated for some of the simulated values of $W_{sp}$ and $\Omega$ for comparison. The perturbations for higher values of $W_{sp}$ is likely due to increased vortex effects, which becomes evident by looking at the animations. Figure \ref{fig:power-comp} (a) and (b) presents the average torque and power generated for varying values of $W_{sp}$ and $\Omega$.  Figure \ref{fig:power-comp} (c) presents the power for varying $\Omega$ against a fixed $W_{sp}$. This allows us to determine the maximum power that can be harvested for a given wind speed - which is of interest to us in context of NREL SAM.

\begin{figure}[htpb]
  \begin{center}
    \subfloat[$C_P$ values obtained for $\Omega = 5$ \& $10$ rad/s]{\includegraphics[scale = 0.45]{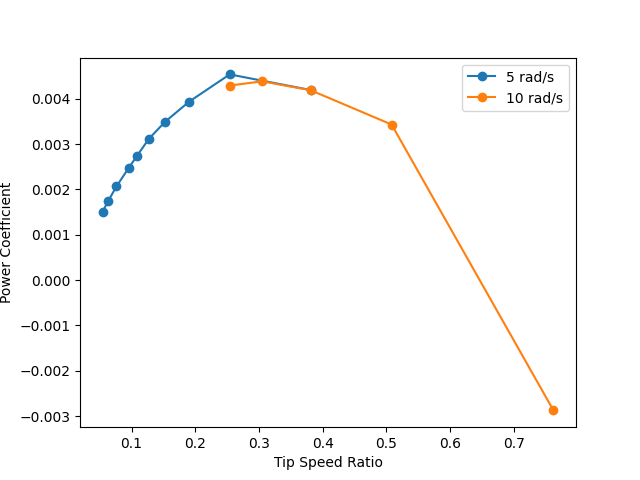}}
    \hfill
    \subfloat[$C_{\tau}$ values obtained for $\Omega = 5$ \& $10$ rad/s]{\includegraphics[scale = 0.45]{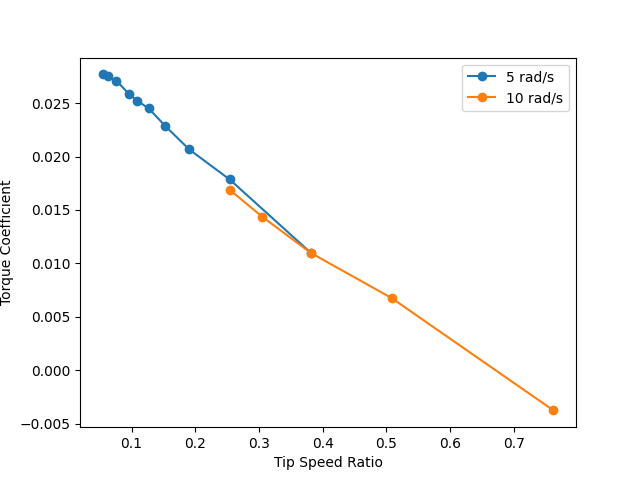}}\\
    \subfloat[Best fit $C_P$ curve obtained]{\includegraphics[scale = 0.45]{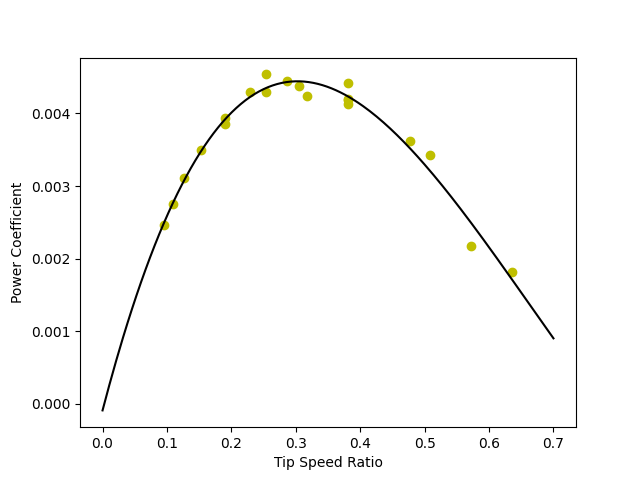}}
    \hfill
    \subfloat[Best fit $C_{\tau}$ curve obtained]{\includegraphics[scale = 0.45]{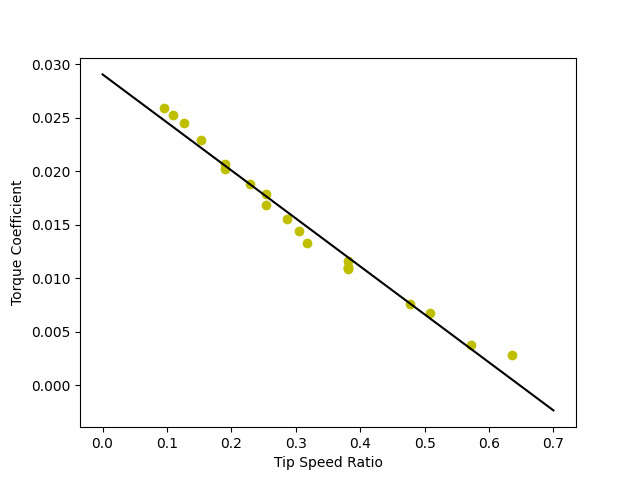}}
    \caption {Power and torque coefficients obtained from simulation results.}
  \label{fig:Cp-Ct}
  \end{center}
\end{figure}

The power and torque coefficients for the turbine can be calculated as:
\begin{equation}
    C_P = \dfrac{\Bar{\tau}\Omega}{\frac{1}{2} \rho S W_{sp}^3} = \dfrac{\Bar{\tau}\Omega}{\rho R H W_{sp}^3}
\end{equation}
\begin{equation}
    C_{\tau} = \dfrac{\Bar{\tau}}{\frac{1}{2} \rho R S W_{sp}^2} = C_P / \lambda
\end{equation}
Figure \ref{fig:Cp-Ct} (a) and (b) present $C_P$ and $C_{\tau}$ curves obtained from simulations for $\Omega = 5$ and $10$rad/s. Figure \ref{fig:Cp-Ct} (c) and (d) present the best fit $C_P$ and $C_{\tau}$ curves obtained using all simulations (including $\Omega \neq 5$ or $10$ rad/s).

Upon comparing with experimental and numerical results of torque and power for VAWT's of similar sizes in [\citenum{dhiman-vawt}], it can be stated that the results are in similar range and the plots retain the characteristics expected from a straight-bladed Darrieus turbine. The $C_P$ and $C_{\tau}$ curves also retain the expected characteristics, but have much smaller magnitudes than [\citenum{dhiman-vawt}]. This anomaly hasn't been resolved yet.

\section{Analysing the Starting Characteristics}

While starting characteristics of a VAWT play no role in energy analysis softwares, it is a useful analysis to have handy - particularly when Darrieus turbines are known to have poorer performance than Savonios and Hybrid turbines at low wind speeds [\citenum{dhiman-vawt}, \citenum{kirke-phd}]. The torque produced by the turbine at low speeds can be determined from the $C_{P/\tau}$ curves generated from the previous section. This section studies the build up of angular velocity, $\Omega$, from a stationary state for varying wind speeds and number of blades.

\begin{figure}[htpb]
  \begin{center}
    \includegraphics[scale = 0.5]{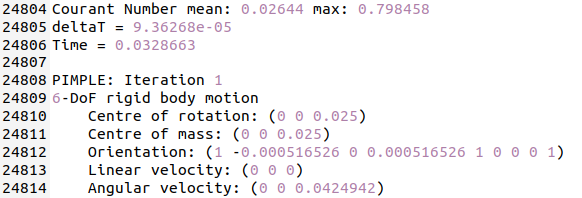}
  \caption {Screenshot of logSolve file indicating angular velocity}
  \label{fig:logSolve}
  \end{center}
\end{figure}

The motion solver \verb+sixDoFRigidBodyMotion+ [\citenum{6dof-wiki}] has been used for the simulations in this section. The solver has been cited in \verb+dynamicMeshDict+ and operates with the VAWT having net $0$ force and moment, with the constraint that motion can only be about (rotation) the z-axis. Thus, the pressure and drag moments are evaluated for the orientation and angular momentum at a given time step, and the next time step's orientation and angular momentum is evaluated taking $\tau_z = 0$. At each time step, the rotor's angular velocity and few other properties are reported onto the terminal, which is stored into the \verb+logSolve+ file (Figure \ref{fig:logSolve}). The time and corresponding angular velocity is extracted from the file, following which it can be plotted.

The motion solver requires inputs about the VAWT including the mass, moment of inertia and centre of mass. The Python library \verb+numpy-stl+ is used for this purpose, which return results for $\rho = 1 $kg/m\textsuperscript{3}. Densities of $1500$ and $3000$ kg/m\textsuperscript{3} have been arbitrarily chosen to determine the solver inputs, which have been tabulated in Table  \ref{tab:6dof-inputs}. 

\begin{table}[htbp]
  \caption{Inputs for sixDoFRigidBodyMotion solver}
  \begin{center}
  \begin{tabular}[c]{|c|c|c|c|c|c|c|} \hline
    S.No. & $N$ & $\rho$ (kg/m\textsuperscript{3}) & Mass (kg) & $I_{xx}$ (kg-m\textsuperscript{2}) & $I_{yy}$ (kg-m\textsuperscript{2}) & $I_{zz}$(kg-m\textsuperscript{2}) \\ \hline
    1 & 2  &  1500 & 0.06871 & 0.00171 & 0.00472 & 0.00641 \\
    2 & 3  &  1500 & 0.10307 & 0.00483 & 0.00483 & 0.00961 \\
    3 & 4  &  1500 & 0.13742 & 0.00644 & 0.00644 & 0.01282 \\
    4 & 3  &  3000 & 0.20613 & 0.00961 & 0.00961 & 0.01932\\\hline
  \end{tabular}
  \label{tab:6dof-inputs}
  \end{center}
\end{table}

\begin{figure}[htpb]
  \begin{center}
    \subfloat[Varying Wind Speed]{\includegraphics[scale = 0.65]{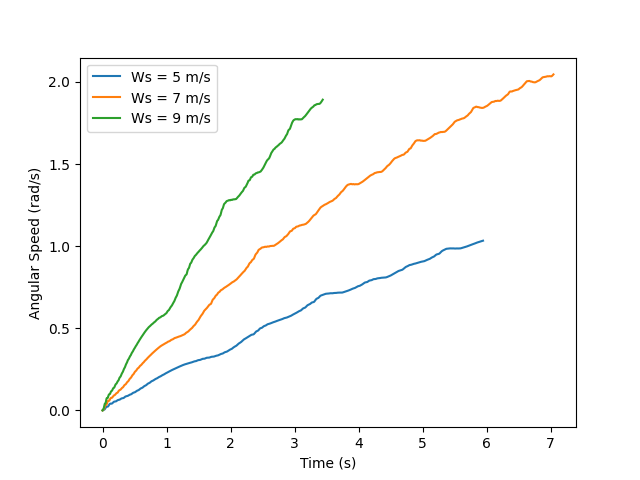}}
    \hfill
    \subfloat[Varying Number of Blades]{\includegraphics[scale = 0.65]{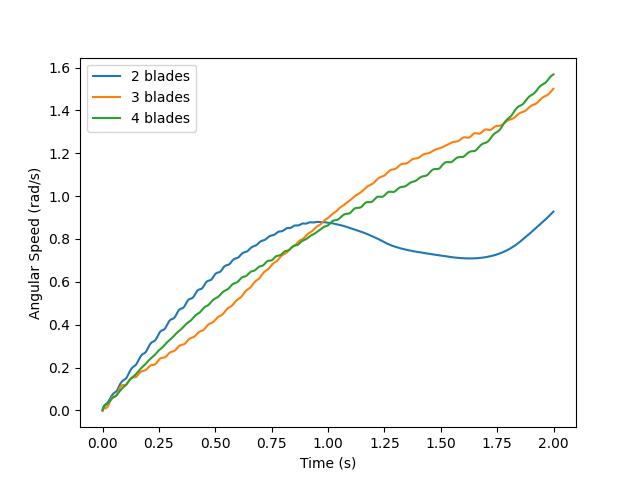}}
    \caption {Angular velocity buildup for varying geometries and wind speeds}
  \label{fig:SS_VAWT}
  \end{center}
\end{figure}

Fig \ref{fig:SS_VAWT} shows a comparison between the angular velocity build up for different cases. \ref{fig:SS_VAWT}(a) compares cases with identical turbines, having properties of S.No 4 in Table \ref{tab:6dof-inputs}, with wind speeds of 5, 7 and 9 m/s. \ref{fig:SS_VAWT}(b) compares turbines with different geometries (S.No 1 - 3 in Table \ref{tab:6dof-inputs}) with identical wind speeds of 7 m/s. Due to the time taken by simulations of  sixDoFRigidBodyMotion solver, none of the graphs were able to reach terminal angular velocity. However, a comparison between the angular accelerations can be seen from the plots. It can also be noted that for $N = 2$ blades, with lower solidity and greater separation between adjacent blades than for $N = 3$ or $4$, the torque produced and the resulting angular acceleration aren't as uniform as for $N = 3$ or $4$.

%%%%%%%%%%%%%%%%%%%%%%%%%%%%%%%%%%%%%%%%%%%%%%%%%%
% Wind Speed Scenario Synthesis.

\chapter{WIND SPEED SCENARIO SYNTHESIS}
\label{chap:arma}

Research concerning weather data analysis can be classified as point forecast and scenario generation. While the former focuses on predicting exact future values for a given period, the latter focuses on generating waveform with identical or similar statistical characteristics. For techno-economic models such as NREL's SAM and REOpt, and HOMER, scenario generation comes in particularly handy for a more detailed analysis. Different synthesis algorithms have been proposed in literature, one of them being the use of auto-regressive moving average (ARMA) to characterize historical weather measurements. \cite{arma-dir} and \cite{arma-FS} are examples of such methods. In [\citenum{arma-dir}], the hourly wind speeds, which resemble a Weibel distribution, is first normalized following which an ARMA model is fit. Using the parameters obtained from the fitting process, new scenarios are generated. In [\citenum{arma-FS}], the seasonality and trend is first removed from the wind speed data before normalizing and fitting the ARMA model. In order to remove seasonality, each month's data is addressed separately, and a Fourier series is used to characterize the seasonal trend. The ``de-trended" data is then used and auto-correlation in the obtained residue is captured. The two studies are used as the reference in this project, and hourly wind speeds are synthesized with an ARMA model using historical data from NOAA\footnote{Since the data from NOAA was chargeable, it was instead obtained from a \href{https://github.com/ColasGael/Machine-Learning-for-Solar-Energy-Prediction}{GitHub project}} [\citenum{noaa}] and TMY2 [\citenum{tmy2}].

\section{Theoretical Background}

An auto-regressive moving average (ARMA) model provides a framework to characterize a stationary (probability density function doesn't vary with time) series by establishing a linear dependence with the past values and a corresponding series of white noise [\citenum{madsen-TSA}]. A generic ARMA ($p,q$) model of a process $\{Z_t\}$ is given by:
\begin{equation}
    Z_{t} = \sum_{i=1}^{p} \phi_{i} Z_{t-i} + \sum_{j=1}^{q} \theta_{j} \epsilon_{t-j} + \epsilon_{t}
\label{eqn:arma-defn}
\end{equation}
where $\phi_i$ and $\theta_j$ are the AR and MA parameters, and $\epsilon_t$ denotes the white noise associated with that time instant. When $p = 0$, the model becomes an MA($q$) model where dependence exists on the previous $q$ noise terms alone. When $q = 0$, the model becomes an AR($p$) model where dependence exists on the previous $p$ values alone, with no dependence on the past noises. HOMER Pro uses an AR($1$) model to synthesize values for daily wind speeds for a user defined $\phi_1$ and white noise standard deviation $\sigma_{\epsilon}$.

\subsubsection*{De-trending and normalizing}

An almost stationary time series can be considered to have 4 components - level, trend, seasonality and noise. While level (average) and noise is what is reflected in an ARMA model, the trend and seasonality are unwanted components. A series without trend and seasonality would be ``flat-looking" and without periodic fluctuations. To arrive at a stationary process, [\citenum{arma-FS}] analyses the wind speed data one month at a time to avoid seasonality, and fits a Fourier series to remove any additional trend and seasonality.

Data obtained from NOAA is for a continuous year and some deviation from a stationary process is observed. Figure 1.1 presents a comparison between plots for 45 days of data with and without de-trending using Fourier series. An improvement in characterization by the model can be seen. Data from TMY2 is, however, not for a continuous year. It instead considers monthly data from $1961 - 1990$ and concatenates weather data that ``best presents" the weather phenomenon of that location. Due to this, the need for de-trending the data like for NOAA isn't there.

While analysing the NOAA data, the trend and seasonality in the wind speed is represented using a Fourier series as presented in Eqn. \ref{eqn:Fourier}.
\begin{equation}
    W_{F}(t) = \sum_{}^k \left[ A_{k} \sin(f_{k} t) + B_{k} \cos(f_{k} t) \right]
\label{eqn:Fourier}
\end{equation}
$\{f_k\}$ is the set of frequencies used, and has been taken as $\{ \frac{1}{500}, \frac{2}{500}, \frac{3}{500}, \frac{4}{500}, \frac{5}{500}, \frac{7}{500} \}$. For finding the optimal $\{ A_k, B_k\}$, least square regression has been implemented - which leads to a set of 12 equations and 12 variables. The rest of the model fitting continues as is with the wind speed residuals, $W_r$($= W_{sp} - W_F$), considered instead of wind speeds.

Finally, to obtain a stationary process, the data is mapped to a normal distribution as described in Eqn. \ref{eqn:norm}
\begin{equation}
    \{Z_{t}\} = \Omega_{N}^{-1} \Omega_{W} \{W_{t}\}
\label{eqn:norm}
\end{equation}
where $\Omega_{W}$ and $\Omega_{N}$ are cumulative distribution functions for $W_{sp}$ or$W_r$ and normal distribution respectively. The CDF for both distributions is evaluated numerically by implementing bins over the range.

\subsubsection*{Fitting an ARMA Model}

The fitting process determines the most suitable $\boldsymbol{\Theta}$ - which represents the parameters and white noise's standard deviation - $\{ \phi_1, \phi_2, ..., \theta_1, \theta_2, ..., \sigma_{\epsilon} \}$. This can be done by minimizing $\sigma_{\epsilon}$, as followed in the project, or by using a Maximum Likelihood Estimator, $MLE$ (see [\citenum{arma-FS}]).

The fitting is done through an iterative process starting with a guess value for $\boldsymbol{\Theta}$. The white noise at each instant, $\epsilon_t$, is estimated using the known process values, ${Z_t}$, and the guess value for the ARMA parameters. For the first $q$ instants, we assume that the process has no white noise. Meaning $\epsilon_i = 0$.
Following it, for $t > q$, we estimate the white noise as:
\begin{equation}
    \epsilon_t = Z_{t} - \left[ \sum_{1}^{p} \phi_{i} Z_{t-i} \right] - \left[ \sum_{1}^{q} \theta_{j} \epsilon_{t-j} \right]
\end{equation}
In the programs, a grid search was implemented to determine the optimal $\boldsymbol{\Theta}$ which leads to the minimum $\sigma_{\epsilon}$.

\subsubsection*{Generating Scenarios}

To generate wind speed scenarios, a set of $\{Z_t\}$ is first generated using $\boldsymbol{\Theta}$ - which is then mapped back to a wind speed or wind speed residual using Eqn \ref{eqn:denorm}.
\begin{equation}
    \{W_{t}\} = \Omega_{W}^{-1} \Omega_{N} \{Z_{t}\}
\label{eqn:denorm}
\end{equation}
While generating $\{Z_t\}$, the first $p$ points are retained as it is with the corresponding $\epsilon_{t} = 0$, and the subsequent $\{ \epsilon_{t}, Z_{t} \}$ are sequentially generated and evaluated respectively using Equations \ref{eqn:arma-defn} and \ref{eqn:whiteN}. To generate the white noise, the Box-Muller transform is used:
\begin{equation}
    \epsilon_{t} =  \sigma_{\epsilon} \times \cos{(2 \pi v_2)} \times \sqrt{-2 \log v_1}
\label{eqn:whiteN}
\end{equation}
where $v_1$ and $v_2$ are random variables in the range [0,1].

\begin{figure}[htpb]
  \begin{center}
    \includegraphics[scale = 0.3]{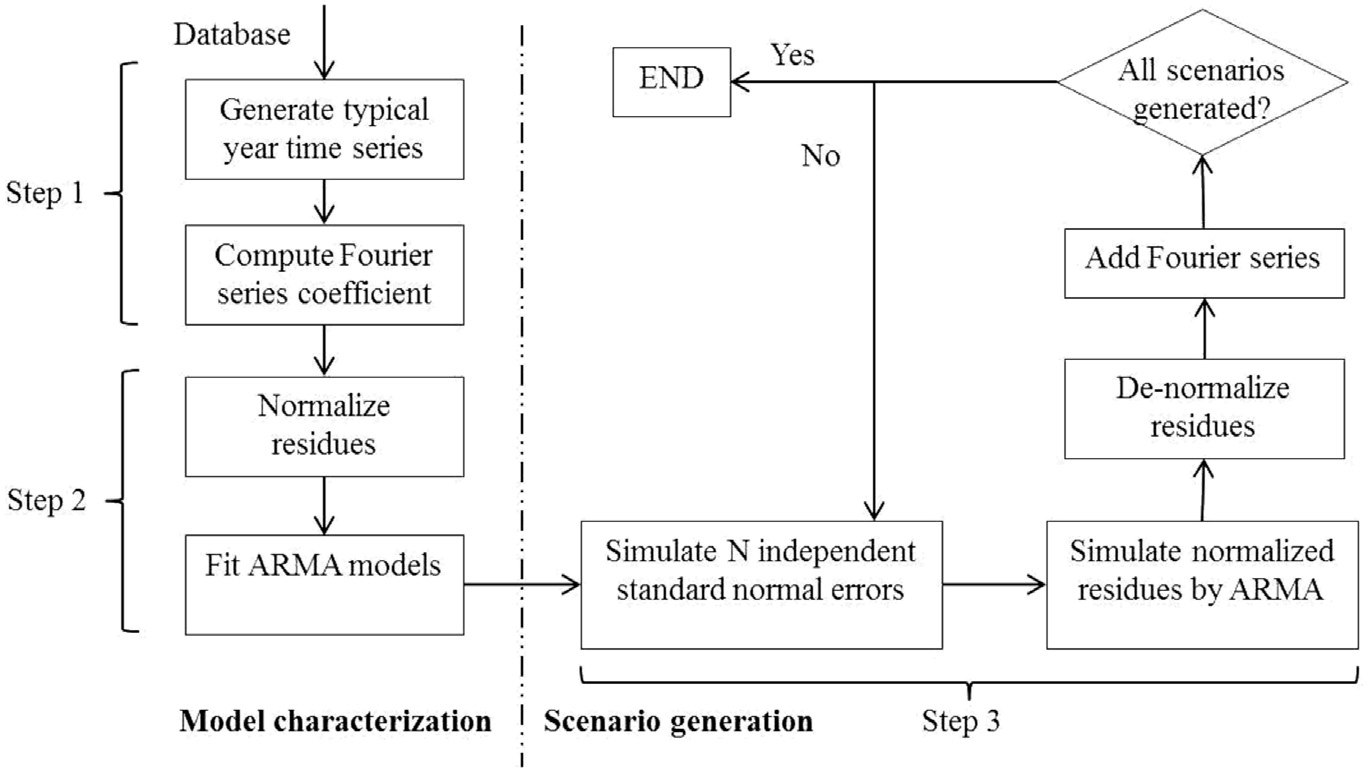}
    \caption {Flow diagram for wind speed synthesis as illustrated by [\citenum{arma-FS}]}
  \label{fig:arma-flow}
  \end{center}
\end{figure}

A flow diagram summarizing the series of steps involved in the generating synthetic wind speeds is presented in Figure \ref{fig:arma-flow}. For models without the need for removing trends and seasonality, ``Computing Fourier series coefficient" in Step 1 and ``Add Fourier series" in Step 3 have to be ignored.

\section{Plots and Discussions}

The model fitting and scenario generation for hourly wind speeds have been programmed in C. Improvement in computational speed through parallel computing using OpenMP has also been explored as part of the course ID5130. The study was for ARMA(1,1) model with Fourier series de-trending, and a negligible improvement for 2 threads was found. More about it can be found in \verb+ID5130Report.pdf+.

\subsubsection*{Impact of de-trending}

From NOAA, hourly weather data, including wind speed, for a particular location from Feb 2016 to Oct 2017 was retrieved. From it, wind speeds across 45 days (1080 hours/points) was chosen for the following analysis.

\begin{figure}[htpb]
  \begin{center}
    \subfloat[Individual comparison between synthetic and historical data]{\includegraphics[scale = 0.43]{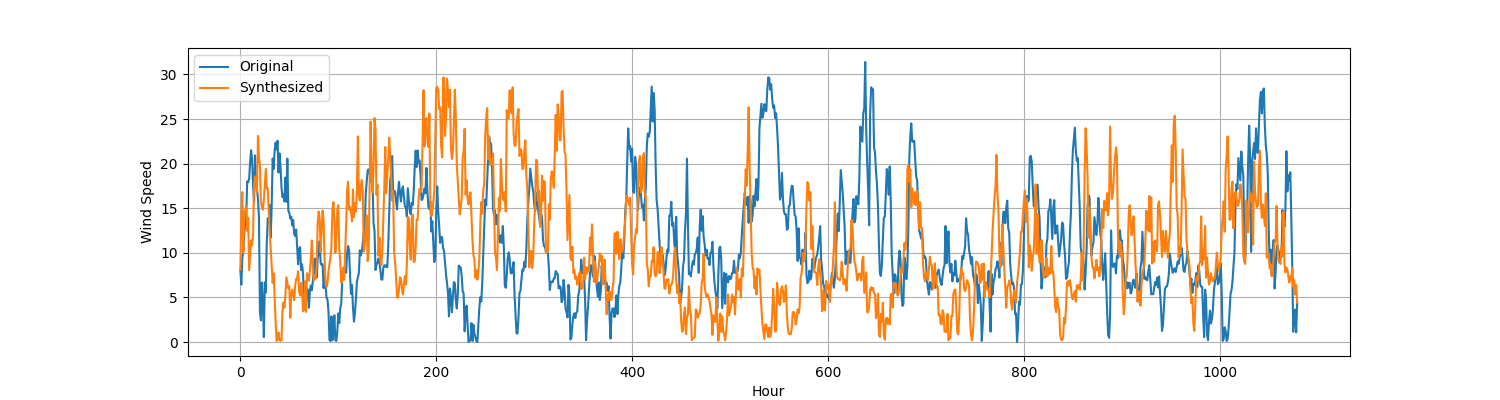}}\\
    \subfloat[6 distinct generated scenarios]{\includegraphics[scale = 0.43]{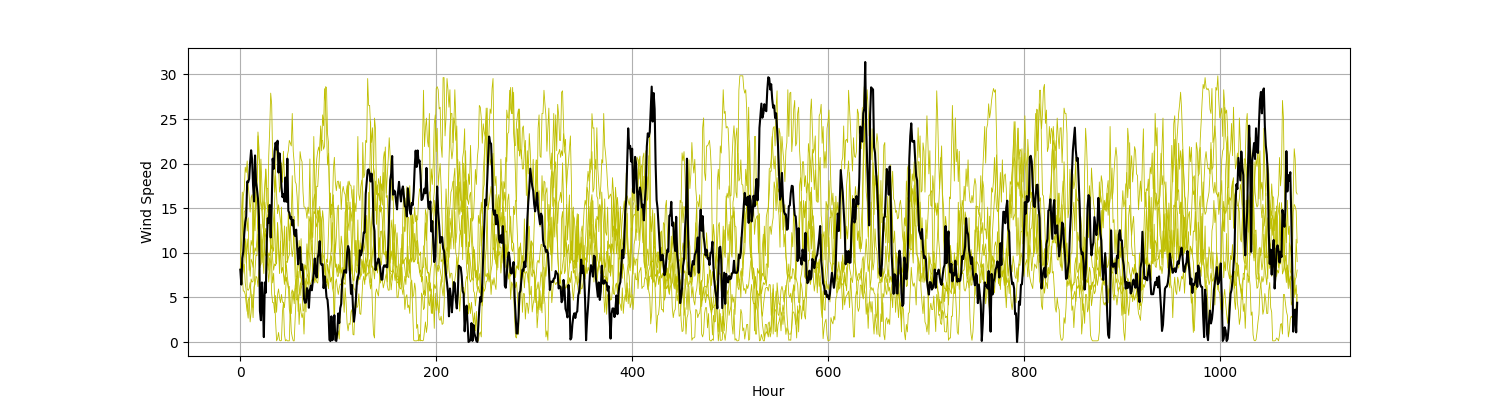}}
    \caption {ARMA(1,1) without de-trending prior to normalization}
  \label{fig:noaa-D}
  \end{center}
\end{figure}

\begin{figure}[htpb]
  \begin{center}
    \subfloat[Individual comparison between synthetic and historical data]{\includegraphics[scale = 0.43]{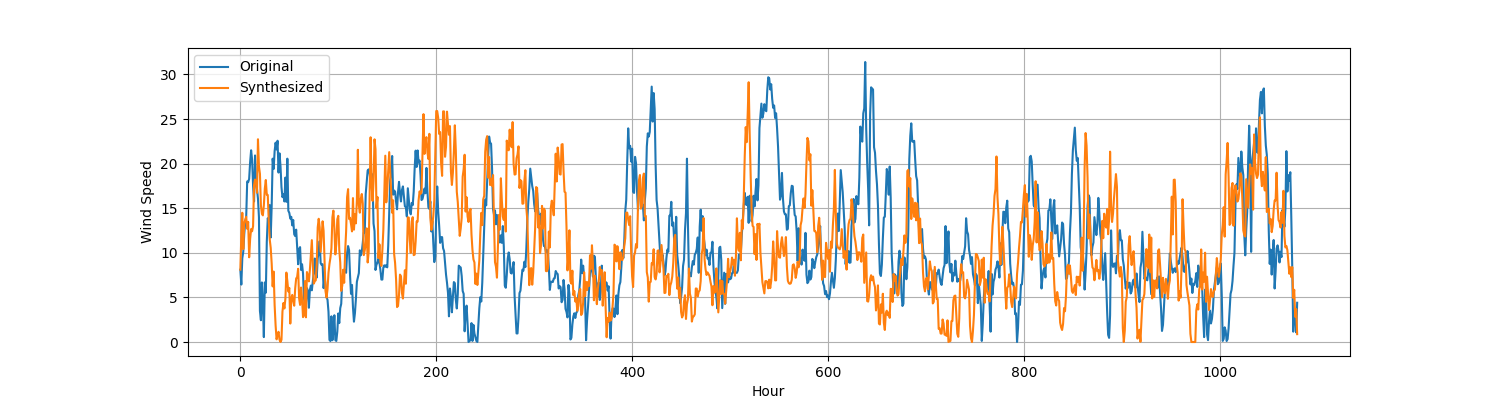}}\\
    \subfloat[6 distinct generated scenarios]{\includegraphics[scale = 0.43]{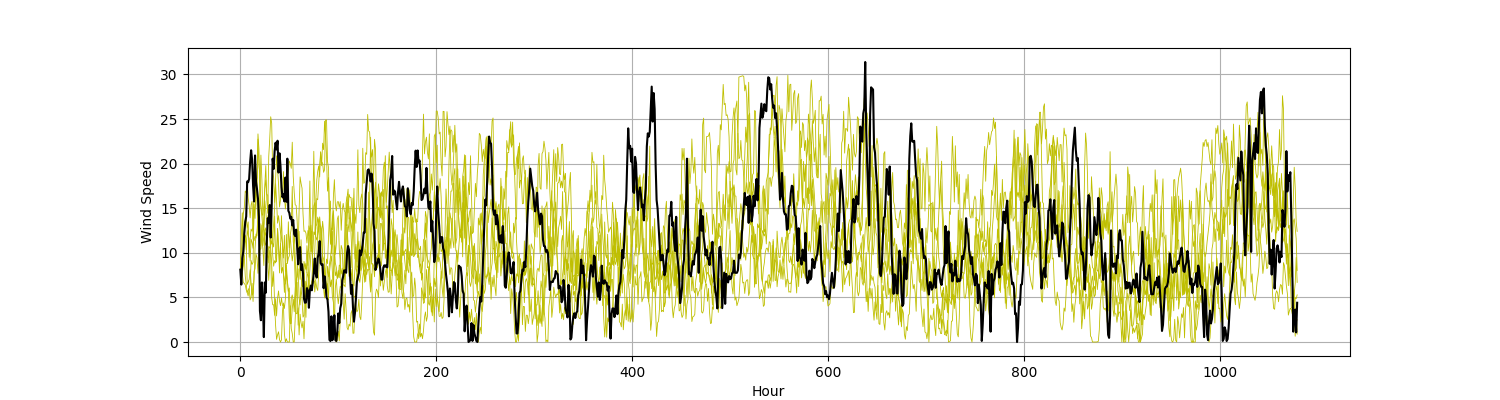}}
    \caption {ARMA(1,1) with de-trending prior to normalization}
  \label{fig:noaa-F}
  \end{center}
\end{figure}

For the scenario generation a total of 6 models have been developed - ARMA (1,1), (1,2) and (2,2), with and without de-trending. The best fit parameter values and obtained statistical characteristics have been summarized in Tables \ref{tab:noaa-p} and \ref{tab:noaa-s}. Figures \ref{fig:noaa-F} and \ref{fig:noaa-D} present a comparison between the plots produced by the two kinds of models. From Table \ref{tab:noaa-s} it can be said that ARMA (1,2) and (2,2) models for this particular use case provided little improvement, and considering the relative fitting simplicity for ARMA (1,1) models, it can be the preferred model. On comparing, it can be said that the seasonality within the month/period is retained to a significant extent in Fig \ref{fig:noaa-F} than Fig \ref{fig:noaa-D}, however, the latter matches with historical statistical characteristics to a greater extent as observed in Table \ref{tab:noaa-s}.

\begin{table}[htbp]
  \caption{Optimum parameters obtained for the 6 models}
  \begin{center}
  \begin{tabular}[]{|c|c|c|c|c|c|c|} \hline
     & Model & $\phi_1$ & $\phi_2$ & $\theta_1$ & $\theta_2$ & $\sigma_{\epsilon}$ \\
    \hline
    \multirow{3}{5em}{W/o de-trending} & ARMA (1,1) & 0.913 & - & 0.078 & - & 0.378\\
     & ARMA (1,2) & 0.915 & - & 0.079 & -0.035 & 0.377\\
     & ARMA (2,2) & 0.910 & 0.005 & 0.100 & -0.035 & 0.377\\
    \hline
    \multirow{3}{5em}{W/ de-trending} & ARMA (1,1) & 0.897 & - & 0.101 & - & 0.399\\
     & ARMA (1,2) & 0.898 & - & 0.104 & -0.028 & 0.397\\
     & ARMA (2,2) & 0.890 & 0.005 & 0.115 & -0.015 & 0.397\\
    \hline
  \end{tabular}
  \label{tab:noaa-p}
  \end{center}
\end{table}

\begin{table}[htbp]
  \caption{Statistical characteristics of the synthesized and obtained wind speeds}
  \begin{center}
  \begin{tabular}[]{|c|c|c|c|c|c|c|c|} \hline
    & Historical & (1,1), w/ & (1,2), w/ & (2,2), w/ & (1,1), w/o & (1,2), w/o & (2,2), w/o \\
    \hline
    Mean & 11.06 & 10.58 & 10.55 & 10.55 & 10.67 & 10.61 & 10.61 \\
    \hline
    Variance & 37.74 & 31.41 & 31.35 & 30.51 & 40.06 & 41.10 & 40.96\\
    \hline
  \end{tabular}
  \label{tab:noaa-s}
  \end{center}
\end{table}

We can thus state that the method is useful for synthesizing wind speed scenarios even when the historical data available doesn't represent a stationary process.

\subsubsection*{Making use of daily variation}

TMY provides weather data that best represents the weather phenomenon for 239 locations. It is the standard data set used by NREL and is made available for analysis in both SAM and HOMER. As stated before, the data obtained from TMY2 was sufficiently close to a stationary process that further modification prior to normalization was found unnecessary.

\begin{figure}[htpb]
  \begin{center}
    \subfloat[Individual comparison between synthetic and historical data]{\includegraphics[scale = 0.43]{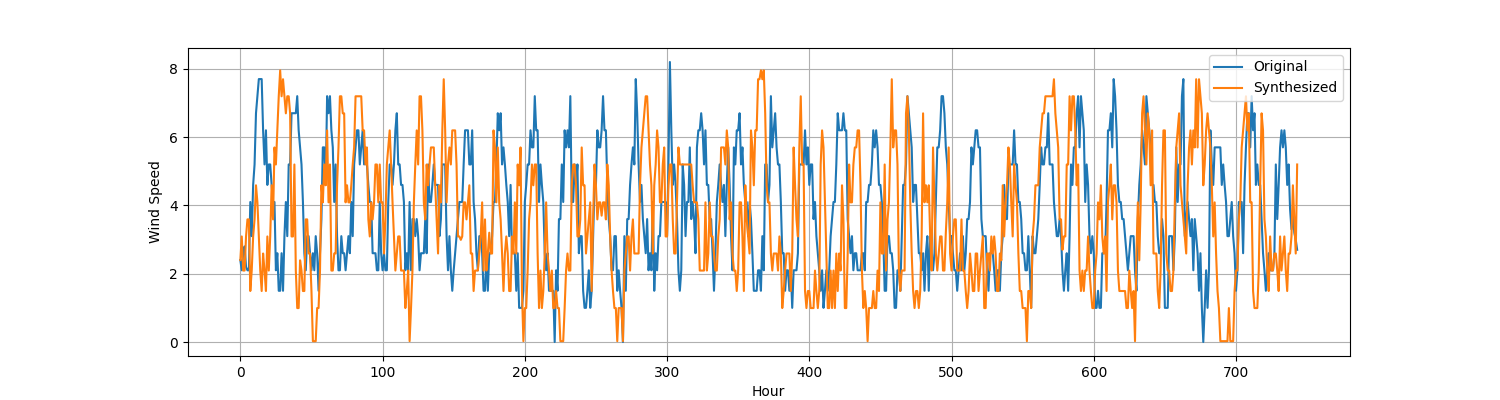}}\\
    \subfloat[6 distinct generated scenarios]{\includegraphics[scale = 0.43]{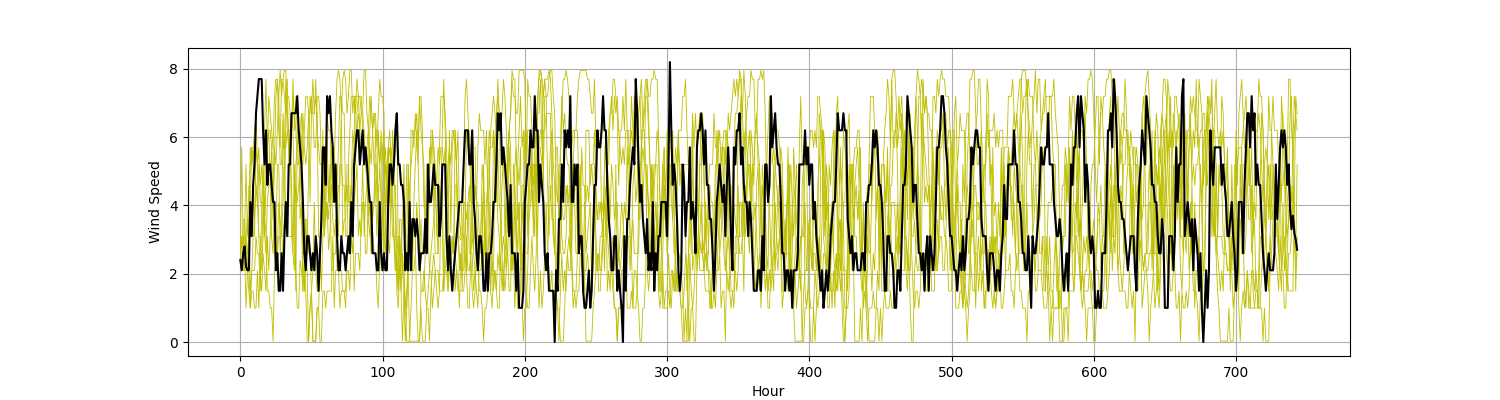}}
    \caption {ARMA(1,1) applied on TMY2 data}
  \label{fig:tmy2-D}
  \end{center}
\end{figure}

From the repository, wind speed for August at Los Angeles and San Diego (regions sufficiently close with similar climates) were extracted. An ARMA(1,1) model was fit to LA's data and scenarios presented in Figure \ref{fig:tmy2-D} were obtained.

\begin{figure}[htpb]
  \begin{center}
    \subfloat[Individual comparison between synthetic and historical data]{\includegraphics[scale = 0.43]{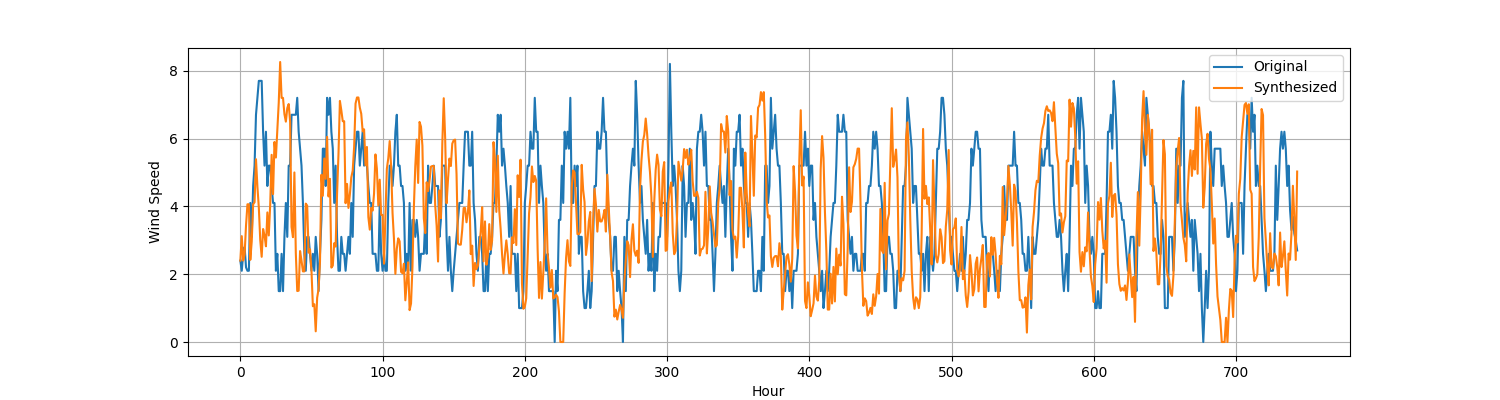}}\\
    \subfloat[6 distinct generated scenarios]{\includegraphics[scale = 0.43]{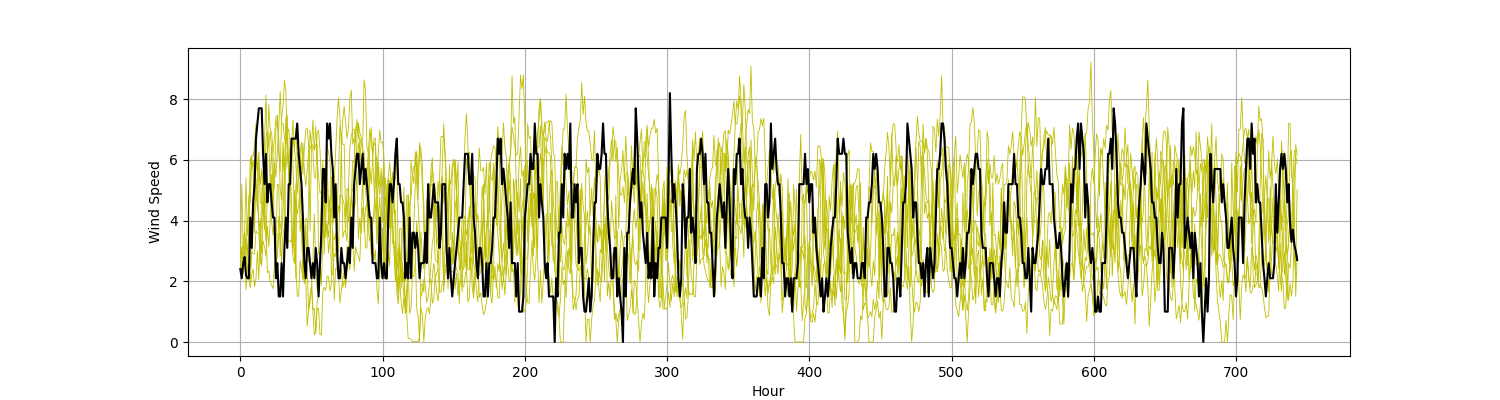}}
    \caption {ARMA(1,1) applied on deviation from daily mean, TMY2 data}
  \label{fig:tmy2-M}
  \end{center}
\end{figure}

A second method has also been implemented, wherein the daily average wind speed has been subtracted and the residuals are used for normalizing and model fitting, like the procedure for NOAA data. The scenarios generated by this method with an ARMA(1,1) model for LA's data has been presented in Figure \ref{fig:tmy2-M}. In this method, Step 1 of Fig. \ref{fig:arma-flow} will have ``Compute daily mean" instead of Fourier series coefficient, and Step 3 will have ``Add daily mean" instead of Fourier series.

\begin{figure}[htpb]
  \begin{center}
    \subfloat[Individual comparison between synthetic and historical data]{\includegraphics[scale = 0.43]{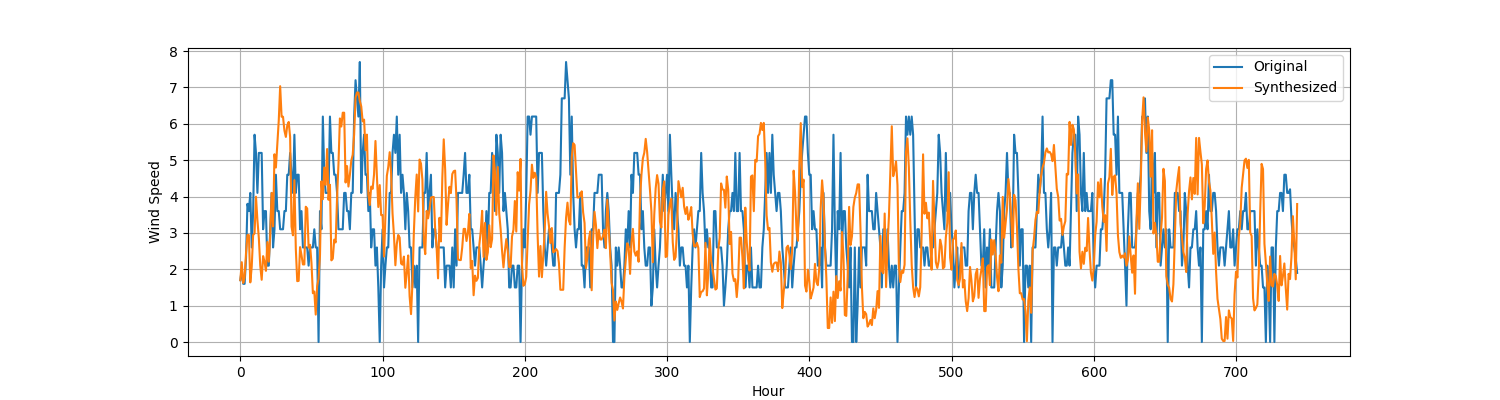}}\\
    \subfloat[6 distinct generated scenarios]{\includegraphics[scale = 0.43]{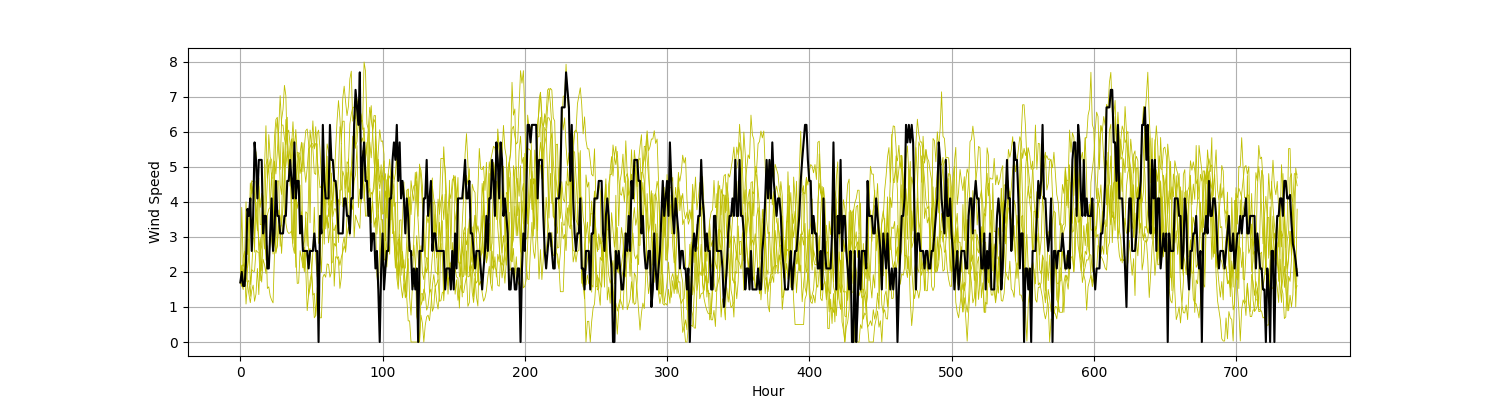}}
    \caption {Hourly wind speed synthesis using daily mean speeds}
  \label{fig:tmy2-B}
  \end{center}
\end{figure}

This method has also been extended to generating wind speed scenarios for San Diego using SD's available daily average wind speeds and LA's deviation from the daily mean. However, with unequal standard deviations for the two locations, a correction/scaling factor ($0.794$) has also been included. Thus the wind speed is evaluated as:
\begin{equation}
    \{W_t\}_{SD} = \{\Bar{W}\}_{SD} + \sigma_{corr}\Omega_{W,LA}^{-1} \Omega_{N} \{Z_{t}\} 
\end{equation}
The scenarios generated using this has been presented in Fig \ref{fig:tmy2-B}, and compared with the actual hourly wind speeds for San Diego.
This method can be particularly useful for running simulations for regions with limited historical weather data. Resources such as NASA POWER [\citenum{nasa-power}] makes it possible to obtain daily average wind speeds for any given location and date. This can be combined with historical hourly data of locations with similar climates, as presented above, to arrive at suitable hourly wind speeds for the given location - which is of interest to softwares such as SAM and HOMER.

%%%%%%%%%%%%%%%%%%%%%%%%%%%%%%%%%%%%%%%%%%%%%%%%%%
% Conclusion.

\chapter{CONCLUSION}
\label{chap:conclusion}

The project was initiated to complement existing techno-economic models such as NREL's System Advisor Model by creating models that address some limitations. With the context of wind power prediction applications, it can be stated that this project tackles the primary limitations a new user comes across. Further, SAM is a cost-free software and has many open-source modules and libraries loaded. Thus, the models/codes were developed to have only open-source dependencies and requiring minimal work by a new user for reproducing results. 

A demonstration video [\href{https://www.youtube.com/watch?v=NbD3zoFkFHA&t=3s}{YT Link}] by NREL for wind power system modelling on SAM is chosen for context for the following statements. In the demonstration, a wind farm has been modelled from scratch using the available libraries, and results such as power generated, investment and payback, and more were obtained. The parameters required for the simulation include wind resource, wind turbine specification, wind farm layout and wake characteristics, system costs and degradation factors, and financial parameters.

Ignoring wake characteristics and using default values for the degradation and financial parameters, a user is concerned with the first two input sections alone: hourly wind speed data and turbine specifications.

Figure \ref{fig:winddata-sam} is a screenshot from the demonstration for wind speed data. A user can either use a CSV file containing hourly wind speed data or provide parameters for a Weibel distribution, the latter of which provides results for average power alone. The library for hourly wind speed data is from the TMY dataset, which is centred around USA. Obtaining hourly weather data outside of it is typically tedious and often a paid service. The paid software ``HOMER Pro" has features for synthesizing wind speed data (using AR(1) model) using user-defined parameters - which are often difficult to estimate. Features for learning from an input existing dataset to generate synthetic scenarios is absent. Chapter \ref{chap:arma} of the report has addressed this limitation. ARMA models have been developed with additional detrending features to characterize existing wind speed data and thus generate new scenarios. ARMA(1,1) models found to perform as good as ARMA(1,2) and (2,2) models. Thus, ARMA (1,1) models were found the most suitable model for wind speed characterization due to their relative simplicity. Two detrending methods - fitting a Fourier series and using daily mean speeds have been explored as well, and the characterization was found to improve. A comparison of the overall statistical characteristics between the synthetic and historical data and graphical comparisons between the two have been presented in the chapter. From it, we can conclude that the synthetic wind speeds sufficiently represents the historical data. The chapter also looks into synthesizing wind speeds for a location using the deviation of hourly wind speeds from daily mean speeds for a location with a similar climate, along with the daily mean speed of that location. This has been implemented for San Diego using the hourly deviation for Los Angeles, a city sufficiently close with similar climates, with TMY2 data. From figure \ref{fig:tmy2-B} it can be said that the synthetic data is sufficiently close to the actual data and is a good representation. With daily mean wind speeds being a much more accessible property from datasets such as NASA POWER, this method is useful for generating data for remote locations. Thus, a new user of SAM with no access to weather data for the location of interest can use the ARMA models developed as an alternative. With the programs written in C, the results are easily reproducible and can also be easily modified.

Figure \ref{fig:turbinedata-sam} is a screenshot from the demonstration for the turbine characteristics. SAM requires the turbine's power curve [Fig. \ref{fig:power-sam-homer}], which tells the maximum power obtained from the turbine for a given wind speed. As stated in chapter \ref{chap:intro}, the SAM libraries contain curves for HAWT's, and power curve tables for VAWT's are absent. With the increasing popularity of VAWT's, particularly in residential use cases, should a new user want to explore its benefits within a particular system, the process can be tedious when simulated on SAM. Chapter \ref{chap:dmsm} and \ref{chap:cfd} demonstrates using analytical and numerical methods to characterize Darrieus turbines, thus generating power curve tables. 

Chapter \ref{chap:dmsm} uses the Double-Multiple Streamtube model for predicting power generated at different wind speeds. The $C_P$ curve obtained for large-sized turbines ($\sim$5 m) was found to retain the characteristics obtained from experimental data within sufficient limits. The model failed to predict ranges of angular speeds and wind speeds with positive net torque for small-sized turbines. A possible reason for this could be erroneous interpolations for $C_L$ and $C_D$ values for high values of local angle of attack. Inconsistencies in the $C_P$ curve for positive blade angle of attack in large-sized turbines presented in Figure \ref{fig:Cp-anlyt-comp} (a) seconds this argument. The model has been coded in Python, retaining nomenclature used in the report, and can be conveniently used by a new user.

Chapter \ref{chap:cfd} uses 2-D OpenFOAM simulations for predicting power generated by a Darrieus turbine at different wind speeds. The power and torque variation and the $C_P$ and $C_{\tau}$ curve were found to retain the shapes/characteristics expected from a Darrieus turbine. However, the predicted values were significantly smaller than what is encountered in literature. OpenFOAM is a cost-free toolbox based in C++, and has a complete text-based interface. Further, the cases are developed with minimal need for editing for varying situations as described in chapter \ref{chap:cfd} section 3. The CFD simulations use STL files to represent the turbine geometry. Codes using the open-source ``numpy-stl" package have been used to generate the turbine geometry from command line within seconds. Additional programs for making the postProcessing convenient have also been developed and included in the uploaded cases. Thus the CFD simulations can be modified and used by a new user with minimal knowledge of CFD, programming or OpenFOAM, thereby generating power curve tables conveniently. However, this comes with the drawback of the ample computational time required for the simulations.

While not relevant to the chosen context of addressing SAM's limitations, CFD simulations for evaluating the starting characteristics of a Darrieus turbine have been developed and presented in chapter \ref{chap:cfd} section 4. 

Thus, with the two sections, this project presents models for characterizing Darrieus turbines and generating synthetic wind speeds, thereby addressing the stated limitations a new user of SAM will have to concern with for wind power system modelling. 

\subsubsection*{Scope for Future Work}

\begin{itemize}
    \item Flow field around a Darrieus turbine is three dimensional in nature [\citenum{dhiman-vawt}], with considerable variation in velocity field across the height of the aerofoils. Thus, developing easy-to-edit 3-D OpenFoam cases will be a desirable feature to be made available for the user. 3-D models will also allow the user to study and compare designs such as curved-bladed and helical-bladed Darrieus turbines. These simulations would however come with the draw-back of requiring far more computational time.
    \item In Chapter \ref{chap:dmsm} it was observed that for high local angles of attack the interpolated values of $C_L$ and $C_D$ were likely erroneous. Programs such as XFOIL [\citenum{xfoil-program}] \footnote{XFOIL wasn't compatible with my OS.} are known to have good interpolation algorithms to determine these coefficients at small Reynolds numbers. Thus, the developed analytical models can be integrated with XFOIL to see if the results improve.
\end{itemize}

%%%%%%%%%%%%%%%%%%%%%%%%%%%%%%%%%%%%%%%%%%%%%%%%%%%%%%%%%%%%
% Appendices.

\appendix

\chapter{CODES}

Codes for the model will be made available upon request.

\noindent
The codes are distributed among the following Folders:
\begin{enumerate}
    \item \verb+Analytical_VAWT+\\
    This contains Python codes for Double-Multiple Streamtube model described in Ch. \ref{chap:dmsm}.
    \item \verb+Numerical_VAWT+\\
    This contains Python and C++ codes for the CFD models described in Ch. \ref{chap:cfd}. The codes are divided into the following sub-folders:
    \begin{enumerate}
        \item \verb+Geometry+\\
        This contains Python codes for creating STL files.
        \item \verb+Const_Omega+\\
        This contains OpenFOAM codes for running CFD simulations for Darrieus turbine running at constant ($W_{sp},\Omega$).
        \item \verb+Start_Char+\\
        This contains OpenFOAM codes for running CFD simulations for Darrieus turbine starting from stationary for a constant $W_{sp}$.
    \end{enumerate}
    \item \verb+Synthetic_Wsp+\\
    This contains C codes for ARMA models  described in Ch. \ref{chap:arma}.
\end{enumerate}
\noindent
\verb+Instructions.pdf+ contains detailed instructions on how to edit and use the above codes for specific use cases.

%%%%%%%%%%%%%%%%%%%%%%%%%%%%%%%%%%%%%%%%%%%%%%%%%%%%%%%%%%%%
% Bibliography.

\begin{singlespace}
  \bibliography{refs}

\begin{thebibliography}{32}
\expandafter\ifx\csname natexlab\endcsname\relax\def\natexlab#1{#1}\fi
\expandafter\ifx\csname url\endcsname\relax
  \def\url#1{{\tt #1}}\fi
\expandafter\ifx\csname urlprefix\endcsname\relax\def\urlprefix{URL }\fi

\bibitem[{Adeyemi(2019)}]{lux-ms}
{\bf Adeyemi, F.} (2019).
\newblock {\em Numerical Simulation of the Lux Vertical Axis Wind Turbine\/}.
\newblock Master's thesis, Department of Mathematics and Statistics, University
  of Saskatchewan.

\bibitem[{Caretto {\em et~al.\/}(1973)Caretto, Gosman, Patankar, and
  Spalding}]{simple-caretto}
{\bf Caretto, L.}, {\bf A.~Gosman}, {\bf S.~Patankar}, and {\bf D.~Spalding},
  Two calculation procedures for steady, three-dimensional flows with
  recirculation.
\newblock {\em In\/} {\em Third International Conference on Numerical Methods
  in Fluid Mechanics\/}. 1973.

\bibitem[{Chen and Rabiti(2017)}]{arma-FS}
{\bf Chen, J.} and {\bf C.~Rabiti} (2017).
\newblock Synthetic wind speed scenarios generation for probabilistic analysis
  of hybrid energy systems.
\newblock {\em Energy\/}, {\bf 120}, 507--517.

\bibitem[{Drela(1989)}]{xfoil-program}
{\bf Drela, M.} (1989).
\newblock Xfoil: An analysis and design system for low reynolds number
  airfoils.
\newblock \urlprefix\url{https://web.mit.edu/drela/Public/web/xfoil/}.

\bibitem[{Hirsch and Mandal(1987)}]{cascade-vawt}
{\bf Hirsch, H.} and {\bf A.~Mandal} (1987).
\newblock A cascade theory for the aerodynamic performance of darrieus wind
  turbines.
\newblock {\em Wind Engineering\/}, {\bf 11}, 164--175.

\bibitem[{{HOMER Energy}(2009--)}]{homer}
{\bf {HOMER Energy}} (2009--).
\newblock {HOMER Pro microgrid software}.
\newblock \urlprefix\url{https://www.homerenergy.com/}.

\bibitem[{Islam {\em et~al.\/}(1994)Islam, Ting, and Fartaj}]{islam-vawt}
{\bf Islam, M.}, {\bf D.~S. Ting}, and {\bf A.~Fartaj} (1994).
\newblock Aerodynamic models for darrieus-type straight-bladed vertical axis
  wind turbines.
\newblock {\em Renewable \& Sustainable Energy Reviews\/}, {\bf 32}.

\bibitem[{Issa(1986)}]{piso-issa}
{\bf Issa, R.} (1986).
\newblock Solution of the implicitly discretised fluid flow equations by
  operator-splitting.
\newblock {\em Journal of Computational Physics\/}, {\bf 62}, 40--65.

\bibitem[{Jacob and Chatterjee(2019)}]{dhiman-vawt}
{\bf Jacob, J.} and {\bf D.~Chatterjee} (2019).
\newblock Design methodology of hybrid turbine towards better extraction of
  wind energy.
\newblock {\em Renewable Energy\/}, {\bf 131}, 625--643.

\bibitem[{Kirke(1998)}]{kirke-phd}
{\bf Kirke, B.~K.} (1998).
\newblock {\em Evaluation of Self-Starting Vertical Axis Wind Turbines for
  Stand-Alone Applications\/}.
\newblock Ph.D. thesis, School of Engineering, Griffith University.

\bibitem[{Lapin(1975)}]{act-discs}
{\bf Lapin, E.~E.} (1975).
\newblock Theoretical performance of vertical-axis wind turbines.
\newblock {\em ASME paper 75-WA/Ener-1\/}.

\bibitem[{Larsen(1975)}]{vortex-vawt}
{\bf Larsen, H.} (1975).
\newblock Summary of a vortex theory for the cyclogyro.
\newblock {\em Proceedings of the second US national conferences on wind
  engineering research\/}.

\bibitem[{Launder and Sharma(1974)}]{turb-k-eps}
{\bf Launder, B.~E.} and {\bf B.~I. Sharma} (1974).
\newblock Application of the energy dissipation model of turbulence to the
  calculation of flow near a spinning disc.
\newblock {\em Letters in Heat and Mass Transfer\/}, {\bf 1}, 131--138.

\bibitem[{Madsen(2008)}]{madsen-TSA}
{\bf Madsen, H.}, {\em Time Series Analysis\/}.
\newblock Chapman \& Hall, 2008.

\bibitem[{Menter(1994)}]{SST-k-om}
{\bf Menter, F.~R.} (1994).
\newblock Two-equation eddy-viscosity turbulence models for engineering
  applications.
\newblock {\em NASA, Ames Research Center; AIAA Journal\/}, {\bf 32}.

\bibitem[{Morales {\em et~al.\/}(2010)Morales, Mínguez, and Conejo}]{arma-dir}
{\bf Morales, J.}, {\bf R.~Mínguez}, and {\bf A.~Conejo} (2010).
\newblock A methodology to generate statistically dependent wind speed
  scenarios.
\newblock {\em Applied Energy\/}, {\bf 87}, 843--855.

\bibitem[{{NASA}(accessed 2021)}]{nasa-power}
{\bf {NASA}} (accessed 2021).
\newblock {Prediction of Worldwide Energy Resources (POWER)}.
\newblock \urlprefix\url{https://power.larc.nasa.gov/}.

\bibitem[{{National Oceanic \& Atmospheric Administration}(accessed
  2017)}]{noaa}
{\bf {National Oceanic \& Atmospheric Administration}} (accessed 2017).
\newblock Hourly weather dataset.
\newblock \urlprefix\url{www.ncdc.noaa.gov/cdo-web/datasets}.

\bibitem[{{National Renewable Energy Laboratory}(2007--)}]{nrel-sam}
{\bf {National Renewable Energy Laboratory}} (2007--).
\newblock {System Advisor Model (SAM)}.
\newblock \urlprefix\url{https://sam.nrel.gov/}.

\bibitem[{{National Renewable Energy Laboratory}(2019)}]{nrel-sam-wind}
{\bf {National Renewable Energy Laboratory}} (2019).
\newblock {Modeling Wind Power Systems in SAM 2018.11.11}.
\newblock \urlprefix\url{https://sam.nrel.gov/wind.html}.

\bibitem[{{National Solar Radiation Database}(1961--1990)}]{tmy2}
{\bf {National Solar Radiation Database}} (1961--1990).
\newblock {Typical Meteorological Year (TMY-2)}.
\newblock \urlprefix\url{https://nsrdb.nrel.gov/about/tmy}.

\bibitem[{{OpenCFD Ltd}(2004--)}]{openfoam}
{\bf {OpenCFD Ltd}} (2004--).
\newblock {Open-source Field Operation And Manipulation}.
\newblock \urlprefix\url{www.openfoam.com}.

\bibitem[{{OpenFOAM Wiki}(accessed 2021{\natexlab{{\em a\/}}})}]{6dof-wiki}
{\bf {OpenFOAM Wiki}} (accessed 2021{\natexlab{{\em a\/}}}).
\newblock {Parameter Definitions - dynamicMotionSolverFvMesh}.
\newblock
  \urlprefix\url{https://openfoamwiki.net/index.php/Parameter_Definitions_-_dynamicMotionSolverFvMesh}.

\bibitem[{{OpenFOAM Wiki}(accessed 2021{\natexlab{{\em b\/}}})}]{simple-wiki}
{\bf {OpenFOAM Wiki}} (accessed 2021{\natexlab{{\em b\/}}}).
\newblock {The SIMPLE algorithm in OpenFOAM}.
\newblock
  \urlprefix\url{https://openfoamwiki.net/index.php/OpenFOAM_guide/The_SIMPLE_algorithm_in_OpenFOAM}.

\bibitem[{Paraschivoiu(1977)}]{dmsm-vawt}
{\bf Paraschivoiu, I.} (1977).
\newblock Double-multiple streamtube model for darrieus wind turbines.
\newblock {\em International Symposium on Wind Energy Systems\/}.

\bibitem[{{Rick van Hattem}(2021)}]{numpy-stl}
{\bf {Rick van Hattem}} (2021).
\newblock {numpy-stl}.
\newblock \urlprefix\url{pypi.org/project/numpy-stl/}.

\bibitem[{Sheldahl {\em et~al.\/}(1980)Sheldahl, Klimas, and
  Feltz}]{sandia-vawt}
{\bf Sheldahl, R.~E.}, {\bf P.~Klimas}, and {\bf L.~Feltz}, Aerodynamic
  performance of a 5-metre-diameter darrieus turbine with extruded aluminum
  naca-0015 blades.
\newblock 1980.

\bibitem[{Sheldahl and Klimas(1981)}]{osti-clcd}
{\bf Sheldahl, R.~E.} and {\bf P.~C. Klimas} (1981).
\newblock Aerodynamic characteristics of seven symmetrical airfoil sections
  through 180-degree angle of attack for use in aerodynamic analysis of
  vertical axis wind turbines.
\newblock {\em Wind Engineering\/}.

\bibitem[{Strickland(1977)}]{strickland-vawt}
{\bf Strickland, J.~H.} (1977).
\newblock A performance prediction model for the darrieus turbine.
\newblock {\em International Symposium on Wind Energy Systems\/}.

\bibitem[{Templin(1974)}]{templin}
{\bf Templin, R.~J.} (1974).
\newblock Aerodynamic performance theory for the nrc vertical-axis wind
  turbine.
\newblock {\em NRC Lab report LTR-LA-190\/}.

\bibitem[{Wilcox(1988)}]{turb-k-om}
{\bf Wilcox, D.~C.} (1988).
\newblock Re-assessment of the scale-determining equation for advanced
  turbulence models.
\newblock {\em AIAA Journal\/}, {\bf 26}, 1299--1310.

\bibitem[{Wilson and Lissaman(1974)}]{wilson-vawt}
{\bf Wilson, R.~E.} and {\bf P.~B.~S. Lissaman} (1974).
\newblock Applied aerodynamics of wind power machines.
\newblock {\em Oregon State University\/}.

\end{thebibliography}
\end{singlespace}

\end{document}